\documentclass[11pt,a4paper]{article}
\usepackage{jheppub} 
\pdfoutput = 1
\usepackage{epsfig}
\usepackage{amsmath,amsfonts,amssymb}
\usepackage{hyperref}
\usepackage{colortbl}
\usepackage{pifont}
\usepackage{subcaption}

\newcommand{\RM}[1]{{\color{blue} [\textbf{RM: } #1] }}

\newcommand{\Ref}[1]{ref.~\cite{#1}}
\newcommand{\fig}[1]{figure~\ref{#1}}
\newcommand{\figs}[2]{figures~\ref{#1} and \ref{#2}}
\newcommand{\Sec}[1]{section~\ref{#1}}
\newcommand{\App}[1]{appendix~\ref{#1}}
\newcommand{\Eq}[1]{eq.~\eqref{#1}}
\newcommand{\Eqs}[1]{eqs.~\eqref{#1}}
\newcommand{\tab}[1]{table~\ref{#1}}

\providecommand{\openone}{\leavevmode\hbox{\small1\kern-3.8pt\normalsize1}}

\newcommand{\ptj}{p_{T\,J}}

\title{A generic anti-QCD jet tagger}

\author[a]{J.~A.~Aguilar--Saavedra,}
\author[b,c]{Jack Collins,}
\author[b,d]{and Rashmish K.~Mishra}

\affiliation[a]{Departamento de F\'{\i}sica Te\'orica y del Cosmos, Universidad de Granada, E-18071 Granada, Spain}
\affiliation[b]{Maryland Center for Fundamental Physics, Department of Physics, University of Maryland, \\College Park, MD 20742, USA}
\affiliation[c]{Department of Physics and Astronomy, Johns Hopkins University, Baltimore, MD 21218, USA}
\affiliation[d]{INFN, Pisa, Italy and Scuola Normale Superiore, Piazza dei Cavalieri 7, 56126 Pisa, Italy}

\abstract{
New particles beyond the Standard Model might be produced with a very high boost, for instance if they result from the decay of a heavier particle. If the former decay hadronically, then their signature is a single massive fat jet which is difficult to separate from QCD backgrounds. Jet substructure and machine learning techniques allow for the discrimination of many specific boosted objects from QCD, but the scope of possibilities is very large, and a suite of dedicated taggers may not be able to cover every possibility --- in addition to making experimental searches cumbersome. In this paper we describe a generic model-independent tagger that is able to discriminate a wide variety of hadronic boosted objects from QCD jets using $N$-subjettiness variables, with a significance improvement varying between 2 and 8. This is in addition to any improvement that might come from a cut on jet mass. Such a tagger can be used in model-independent searches for new physics yielding fat jets. We also show how such a tagger can be applied to signatures over a wide range of jet masses without sculpting the background distributions, allowing to search for new physics as bumps on jet mass distributions.
}

\begin{document}

\maketitle
\flushbottom

\section{Introduction}

Jet tagging algorithms have become an essential tool to explore the high energy frontier at the Large Hadron Collider (LHC). New physics processes are expected to involve the production of highly-boosted top quarks, $W/Z$, and Higgs bosons, which in their hadronic decay give rise to a `fat jet' $J$ where the decay products are highly collimated. In order to distinguish such jets from the background jets resulting from quarks and gluons, generically denoted as `QCD jets', several jet substructure analysis techniques have been developed~\cite{Butterworth:2008iy,Thaler:2008ju,Kaplan:2008ie,Almeida:2008yp, Plehn:2009rk,Plehn:2010st,Thaler:2010tr,Hook:2011cq,Jankowiak:2011qa,Thaler:2011gf,Larkoski:2013eya,Moult:2016cvt,Datta:2017rhs}. These are extensively used in searches for $W' \to t b$~\cite{Aad:2014xra,Sirunyan:2017ukk}, $t \bar t$ resonances~\cite{Sirunyan:2017uhk}, diboson resonances~\cite{Khachatryan:2016cfa,Aaboud:2016okv,Sirunyan:2016cao,Sirunyan:2017wto}, vector-like quarks~\cite{Khachatryan:2015gza,Khachatryan:2016vph}, dark matter produced in association with gauge bosons~\cite{Khachatryan:2016mdm,Aaboud:2016qgg} and new light bosons~\cite{Sirunyan:2017nvi}.

Experimental analyses carried out by the ATLAS and CMS collaborations use dedicated taggers in addition to the jet mass, to search for beyond the Standard Model (BSM) scenarios that can give rise to boosted top quarks, $W/Z$, or Higgs bosons. For example, shape variables such as the $N$-subjettiness ratio $\tau_{21}^{(1)}$~\cite{Thaler:2010tr} and the energy correlation function $D_2^{(\beta = 1)}$~\cite{Larkoski:2013eya} are very effective in distinguishing between QCD jets and two-pronged decays from $W/Z$, and the performance can be further improved by using a more complete set of jet substructure variables and a multivariate analysis~\cite{Datta:2017rhs}. As another example, the subjettiness ratio $\tau_{32}^{(1)}$ is used to identify jets from top quark decays. However, the inherent drawback in this approach is that, while these dedicated taggers are efficient in the discrimination of top quarks and $W/Z$ hadronic decays from QCD jets, they may not be able to identify fat jets arising from the decay of BSM boosted particles.

New particles near the electroweak scale may exist and evade direct detection, for example, if their couplings to quarks and gauge bosons are small. They can still be produced in the decay of heavier particles and may have dominant decays into hadronic final states. Examples of such cases are neutral (pseudo-)scalars in models with left-right symmetry~\cite{Aguilar-Saavedra:2015iew} and warped extra dimensional models with more than 2 branes~\cite{Agashe:2016rle,Agashe:2016kfr} (see also refs.~\cite{Mohapatra:2013cia,Ellwanger:2017skc}). An explicit example of the limitations of dedicated jet taggers has been given in \Ref{Aguilar-Saavedra:2017zuc}, by considering a new `stealth boson' $S$ with a mass in the 100 GeV range and undergoing a cascade decay $S \to AA \to b \bar b b \bar b$ mediated by a lighter particle $A$. When $S$ is boosted, so that the four $b$ quarks merge into a single jet, the $\tau_{21}^{(1)}$ and $D_2^{(\beta = 1)}$ variables used to tag massive SM bosons would `see' the resulting four-pronged fat jet as a QCD jet. Consequently, new physics searches involving boosted hadronically-decaying $W$ or $Z$ bosons, e.g. diboson resonance searches, can be relatively blind to the analogous new physics processes (diboson-like resonances) involving one or two $S$ particles of a mass around $M_{W,Z}$

It is highly desirable that ATLAS and CMS searches are not restricted to a few simple benchmark models, but rather cover as many new physics signatures as possible. A broader scope for LHC searches becomes of the utmost importance given the absence of any convincing hint of new physics beyond the SM,  as we still do not know how new physics may manifest at collider experiments. 
 With that purpose, a generic `anti-QCD' tagger that distinguishes QCD jets not only from $W/Z$ hadronic decays, but also from generic BSM boosted objects, would be a useful tool. In this paper we address this problem and provide a proof of concept that this kind of tool can be developed (see also \Ref{Chakraborty:2017mbz} which pursues related ideas). With this goal, we perform a multivariate analysis using a neural network (NN) that is trained to discriminate QCD jets from fat jets with two-, three- and four-pronged structure, arising from the decay of relatively light boosted particles. After describing our framework in \Sec{sec:2}, we perform a simple analysis in \Sec{sec:3}, to demonstrate the discrimination power for several examples of fat jets from boosted new particles against QCD jets. A comparison between the performance of generic and dedicated taggers is given in \Sec{app:dedicatedvsgeneric}. The decorrelation between the background rejection with tagging based on jet substructure and the jet mass requires a slightly more sophisticated analysis, which is presented in \Sec{sec:4}. Our results are discussed in \Sec{sec:5}. 
Some appendices are devoted to additional details of our analysis. 
In appendix~\ref{app:infoinjet} we study the dependence of the results on the number of input variables for the NNs. In 
appendices~\ref{app:bg-quark-trainOrNot} and \ref{app:bkgComp} we discuss how the results change when we modify the signal flavour composition, and the quark/gluon background composition, respectively.
The dependence of the results on the NN architecture is explored in appendix~\ref{app:arch}. 
In appendix~\ref{app:learning} we examine the issue of whether the taggers only learn jet shapes or they also learn about different signal and background kinematics.  A related issue is the dependence of the results on the specific model for hadronisation and showering; this is addressed in appendix~\ref{app:pythia-herwig}, where we compare the results using two Monte Carlo simulation codes. Finally, in appendix~\ref{app:3prongcolour} we study for completeness the signals of light coloured boosted objects.


\section{Framework}
\label{sec:2}
\RM{}
Following \Ref{Datta:2017rhs}, we characterise the jet substructure by a set of generalised $N$-sub\-jet\-tiness~\cite{Thaler:2011gf} variables
 \begin{equation}
 \tau_N^{(\beta)} = \frac{1}{\ptj} \sum_{i} p_{Ti} \; \text{min} \left\{ \Delta R_{1i}^\beta, \Delta R_{2i}^\beta, \dots, \Delta R_{Ni}^\beta \right\} \,,
 \label{ec:tauN}
 \end{equation}
 with $i$ labelling the particles in the jet, $p_{Ti}$ their transverse momenta, $\Delta R_{Ki} $ their lego-plot distance to the axis $K=1,\dots,N$ and $\ptj$ the jet transverse momentum. As in \Ref{Datta:2017rhs}, in the computation of these variables we use the axes defined by exclusive $k_T$ algorithm~\cite{Catani:1993hr,Ellis:1993tq} with standard $E$-scheme recombination~\cite{Blazey:2000qt}. Ref.~\cite{Datta:2017rhs} proposed the following basis of observables,
 \begin{equation}
 \left\{ \tau_1^{(1/2)}, \tau_1^{(1)}, \tau_1^{(2)}, \dots , \tau_{M-2}^{(1/2)}, \tau_{M-2}^{(1)}, \tau_{M-2}^{(2)}, \tau_{M-1}^{(1)}, \tau_{M-1}^{(2)} \right\} \,,
 \label{ec:taulist}
 \end{equation}
motivated by the requirement to be able to fully reconstruct the $(3M - 4)$-dimensional phase space for a decay into $M$ particles. They found that the discriminating power for $Z$-jets versus gluon and quark jets was saturated by considering up to 4-body phase space. Because we are interested also in higher pronged decays we explore a larger 17-dimensional basis with $M=7$. This specific choice is motivated in \App{app:infoinjet}. It is likely that a smaller, more carefully selected basis of substructure variables could be used with little degradation in discrimination power, but in this work we do not attempt to optimise this. We remark that, equivalently, a set of energy correlation functions~\cite{Larkoski:2013eya} $\text{ECF}(N,\beta)$ could also be used, but the calculations are much more computationally-demanding when one considers higher $N$, as is required for the identification of multi-pronged boosted jets.

The values of these variables are used as the input to a NN trained to discriminate quark/gluon jets from multi-pronged decays of boosted colour singlet particles. Quark and gluon jets are obtained by generating the parton-level processes $p p \to Zg$ and $p p \to Zq$, with $Z \to \nu \nu$, using {\scshape MadGraph5}~\cite{Alwall:2014hca}. Event generation is followed by hadronisation and parton showering with {\scshape Pythia~8}~\cite{Sjostrand:2007gs}. The detector response is simulated with {\scshape Delphes 3.4}~\cite{deFavereau:2013fsa} using the CMS detector card. Jets are reconstructed using the anti-$k_T$ algorithm~\cite{Cacciari:2008gp}  with radius $R=0.8$, as implemented in {\scshape FastJet 3.2}~\cite{Cacciari:2011ma}. For the signal we use fat jets resulting from the decay of neutral, colour-singlet particles into two, three and four quarks. For these, we consider the six processes
\begin{align}
& p p \to Z' \to S \, Z(\to \nu \nu) \,, && S \to u \bar u ~~ \text{ and } ~~ S \to b \bar b , \notag \\
& p p \to Z' \to F \, Z ( \to \nu \nu) \nu \,, && F \to u d d ~~ \text{ and } ~~ F \to u b b, \notag \\
& p p \to Z' \to S \, Z(\to \nu \nu) \,, && S \to u \bar u u \bar u ~~ \text{ and } ~~ S \to b \bar{b} b \bar{b},
\label{ec:MIdata}
\end{align}
with $S$ a scalar and $F$ a fermion. These processes are generated with {\scshape Protos}~\cite{protos} and, in order to remain as model-agnostic as possible, we implement decays of $S$ and $F$ with a flat matrix element, so that the decay weight of the different kinematical configurations only corresponds to the two-, three- or four-body phase space. We will refer to these Monte Carlo data as Model Independent (MI) data in the following.
Our choice is motivated by the need to sample phase space without model prejudice. For example, any specific choice of four-body decay topology, such as $1 \to 1+1 \to 2+2$, combined with a choice of masses for the intermediate particles, would only sample a part of four-body phase space, which varies with those mass choices. Therefore, training on specific cascade modes would introduce a model bias. Our choice to train on both light and $b$ quarks is also with the same aim, of making the tagger as model-agnostic as possible. Variations on this choice, either removing final states with $b$ quarks, or adding signal processes with gluons (e.g. $S \to gg$) in the training, are discussed in appendix~\ref{app:bg-quark-trainOrNot}.

Several new physics signal processes are generated to test whether the NN correctly identifies jets resulting from boosted multi-pronged particle decays, including some for which it is not trained. We use seven such processes,
\begin{align}
& p p \to W' \to W \, Z (\to \nu \nu) \,, && W  \to q \bar q'  \,, \notag \\
& p p \to Z' \to H_1^0 \, Z (\to \nu \nu) \,, && H_1^0  \to gg  \,, \notag \\
& p p \to Z' \to H_1^0 \, Z (\to \nu \nu) \,, && H_1^0  \to A^0 A^0 \,, \notag \\  \displaybreak 
& p p \to Z' \to H_1^0 \, Z (\to \nu \nu) \,, && H_1^0 \to t  \bar t \,, \notag \\
& p p \to Z' \to H_1^0 \, Z (\to \nu \nu) \,, && H_1^0 \to W^+ W^- \,, \notag \\
& p p \to Z' \to H_1^0 \, Z (\to \nu \nu) \,, && H_1^0 \to Z A^0 \,, \notag \\
& p p \to W' \to H^\pm \, Z (\to \nu \nu) \,, && H^\pm \to t b \,,
\label{ec:signals}
\end{align}
with $H_1^0$ a heavy scalar and $A^0$ a pseudo-scalar, $H^\pm$ a charged scalar and $Z'$, $W'$ additional vector bosons. All these new particles arise, for example, in left-right models. We consider hadronic decays of the top quarks, $W/Z$ bosons and pseudo-scalars resulting from the $H_1^0$ and $H^\pm$ decays.\footnote{When the $W$ and $Z$ bosons decay leptonically, the resulting jets have energetic leptons that can be further used for background rejection \cite{Brust:2014gia}. Therefore, we restrict ourselves to the most difficult scenario of fully hadronic decays.}
We note that for $W$ and $Z$ hadronic decays the jet shapes are very similar, so for brevity we only consider the former. These processes are generated with  {\scshape MadGraph5} implementing the relevant interactions~\cite{Aguilar-Saavedra:2015iew} in {\scshape FeynRules}~\cite{Alloul:2013bka}, and using the universal Feynrules output~\cite{Degrande:2011ua} to interface with the event generator.

We treat the search for boosted BSM objects as a binary classification problem, with quark and gluon jets labelled as background and jets originating from boosted massive objects labelled as signal. Our NN classifiers are multilayer perceptrons, a simple fully connected architecture that is well suited for use with unstructured input data. These are implemented using Keras \cite{chollet2015} with a TensorFlow backend \cite{tensorflow2015}. We choose an architecture with two hidden layers, the first containing 512 nodes and the second containing 32 nodes, all using rectifier activation functions. (See appendix~\ref{app:arch} for a few examples using alternative NN architectures.) The output layer is a single node with sigmoid activation. The input consists of the 17 $ \tau_N^{(\beta)}$ variables, with some preprocessing applied. We use two kinds of preprocessing which we discuss in more detail in \Sec{sec:3} and \Sec{sec:4} respectively. The first is a simple standardisation of the inputs, which we find significantly improves the training time, stability over variations of the initial seed, and discrimination performance of the trained tagger for simple architectures like the ones used here. The second approach relies on a more complicated transformation of the input data and allows a tagger to be sensitive to signals over a broad range of masses, while decorrelating background rejection from jet mass and $p_T$.

Except where otherwise specified, we train on equal numbers of background and signal events. Background is divided equally between quark and gluon jets (see appendix~\ref{app:bkgComp} for variations in this choice). Signal training data is divided equally between the six categories of MI events described in \Eqs{ec:MIdata}. 20\% of this signal and background data is set aside for validation. We choose binary cross entropy as the loss function to be optimised, using the Root Mean Square Propagation (RMSProp) algorithm with a learning rate of $10^{-3}$. Additionally, if the loss as measured on validation data does not improve over three epochs, the learning rate is reduced by a factor of 10. Training is stopped after 100 epochs, or when validation loss has not improved in five epochs. Typically, we find that training in this manner takes several tens of epochs. We train five different copies of each NN in the same way but with different starting seeds, and pick the one which has the best performance as measured by area under the receiver operating characteristic (ROC) curve with validation data. All machine learning calculations were performed with an Intel(R) Core(TM) i5-6300U CPU @ 2.40GHz with 8GB of RAM. Our NNs take 1-10 minutes to train.

\section{A first approach to anti-QCD tagging}
\label{sec:3}

For this first simple analysis each of the $N$-subjettiness variables in \Eq{ec:tauN} is standardised by a linear transformation,
\begin{equation}
 \tau_N^{(\beta)} \to  \tau_N^{\text{std} (\beta)} = a_N^{(\beta)} + b_N^{(\beta)} \,  \tau_N^{(\beta)} \,,
 \label{ec:tauNt}
\end{equation}
with $a_N^{(\beta)}$ and $b_N^{(\beta)}$ constant, so that the resulting $\tau_N^{\text{std} (\beta)}$ distribution for the QCD background (composed of equal parts of quark and gluon jets) has zero mean and unit standard deviation. We consider three benchmarks for the jet transverse momentum $\ptj$ and mass $m_J$, and for each one we also select the $Z'$ resonance mass in \Eqs{ec:MIdata} to yield MI data with a  $\ptj$ distribution close to the threshold and similar to the background (see appendix~\ref{app:learning}). One tagger is built in each case,
\begin{itemize}
\item[(a)] Tagger `{\tt std500}': $\ptj > 500$ GeV, $m_J \in [65-105]$ GeV, $M_{Z'} = 1100$ GeV.
\item[(b)] Tagger `{\tt std1000}': $\ptj > 1000$ GeV, $m_J \in [65-105]$ GeV,  $M_{Z'} = 2200$ GeV.
\item[(c)] Tagger `{\tt  std1500}': $\ptj > 1500$ GeV, $m_J \in [350-450]$ GeV,  $M_{Z'} = 3300$ GeV.
\end{itemize}
In the first two cases, the MI data used for training is generated with boosted particle masses $M_{S,F} = 80$ GeV, and in the last case with masses of $400 \; \text{GeV}$. The jet mass $m_J$ and transverse momentum $\ptj$ used here is of ungroomed jet, for reasons that we specify at the end of this section. The number of events used for the training is collected in \tab{tab:stdsamplesizes}.

\begin{table}[htb]
\centering
\begin{tabular}{l|l l l}
						& {\tt std500} & {\tt std1000} & {\tt std1500}\\ \hline
Training sample size 	& 284,016       & 249,206	         & 144,884\\
Validation sample size & 71,004         & 62,302          & 36,220
\end{tabular}
\caption{Training sample sizes for {\tt std} taggers.}
\label{tab:stdsamplesizes}
\end{table}

\begin{figure}[tb]
\begin{center}
\begin{tabular}{cc}
\includegraphics[height=5.8cm]{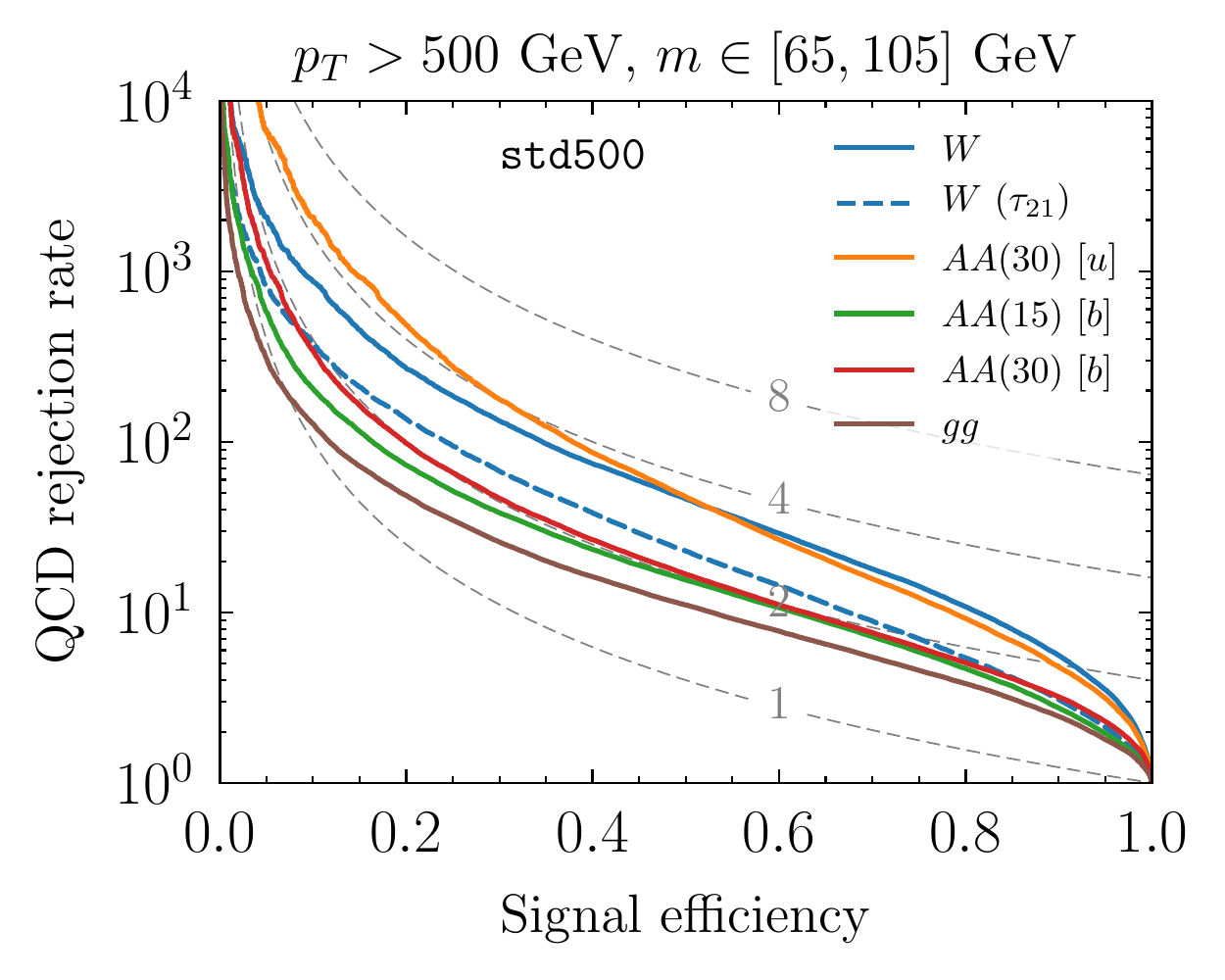} &
\includegraphics[height=5.8cm]{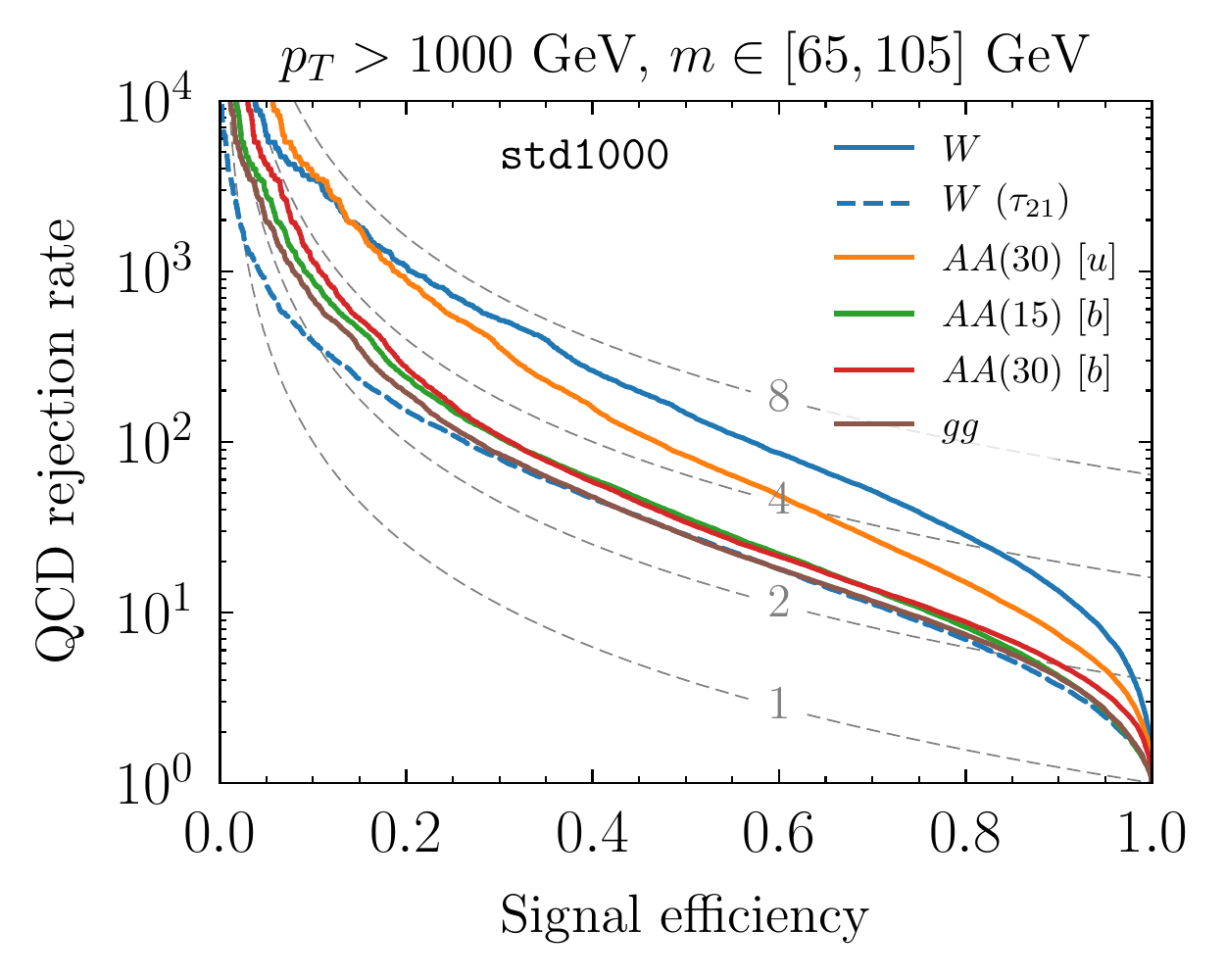}
\end{tabular}
\caption{Signal efficiency versus background rejection for the anti-QCD {\tt std500} (left) and {\tt std1000} (right) taggers. Significance improvement due to shape-variable tagging is indicated by the dashed grey contours. Also indicated is the efficiency curve for $\tau_{21}^{(1)}$ for hadronic $W$ decay (dashed blue).}
\label{fig:500_1000_final}
\end{center}
\end{figure}

\begin{figure}[h]
\begin{center}
\includegraphics[height=5.8cm]{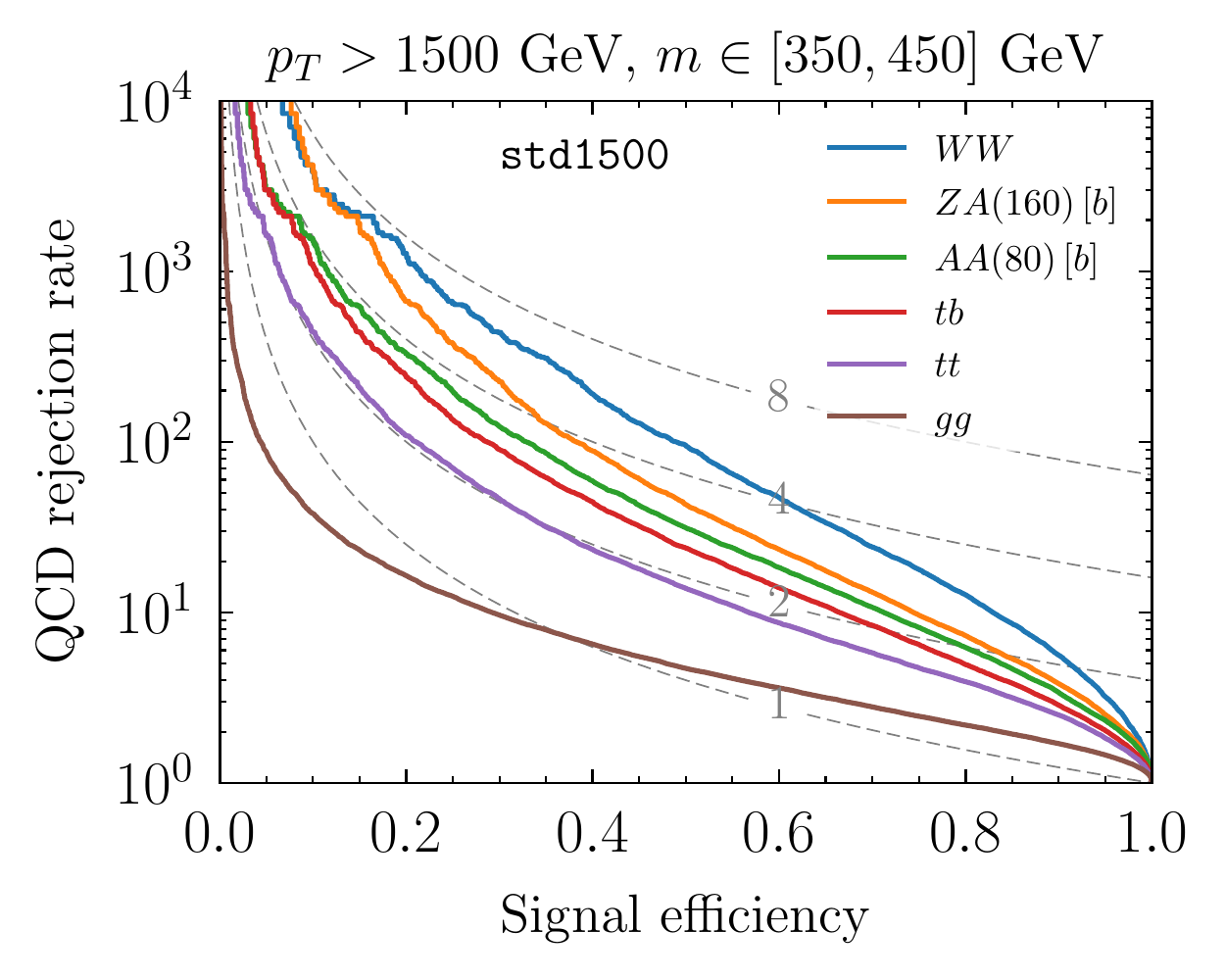}
\caption{Signal efficiency versus background rejection for the  {\tt std1500}  anti-QCD tagger. Significance improvement due to shape-variable tagging is indicated by the dashed grey contours. \label{fig:1500_final}}
\end{center}
\end{figure}

The {\tt std500} and {\tt std1000} taggers are used to investigate the discrimination power for jets coming from BSM boosted particles with a mass around the $W$ mass. We use two regimes of $\ptj$ ($p_T > 500 \; \text{GeV}$ and $p_T > 1000 \; \text{GeV}$) to check the differences, and the extent to which the results are specific --- or not --- to a kinematical region.
For each benchmark, a NN is trained and validated with MI and QCD data, and is then tested on a number of boosted jet topologies,\footnote{We consider here colour-singlet new particles, as they are the most likely ones that could be mainly seen from the decay of a heavier one, and not from their direct production. An example of the application of the tagger to a coloured particle is presented in appendix~\ref{app:3prongcolour}.}
\begin{align}
& W \to q \bar q' \,, \notag \\
& H_1^0 \to gg  \,, && M_{H_1^0} = 80~\text{GeV} \,, \notag  \\
& H_1^0 \to A^0 A^0 \to b \bar b b \bar b \,,  &&M_{H_1^0} = 80~\text{GeV} \,, \quad M_{A^0} = 30~\text{GeV} \,, \notag \\
& H_1^0 \to A^0 A^0 \to b \bar b b \bar b \,,  && M_{H_1^0} = 80~\text{GeV} \,, \quad M_{A^0} = 15~\text{GeV} \,, \notag \\
& H_1^0 \to A^0 A^0 \to u \bar u u \bar u \,, && M_{H_1^0} = 80~\text{GeV} \,, \quad M_{A^0} = 30~\text{GeV} \,, \label{eq:80gevprocs}
\end{align}
setting the parent $Z'$ resonance mass responsible for the processes in \Eq{eq:80gevprocs} to 1100 GeV and 2200 GeV for the $p_T > 500 \; \text{GeV}$ and $p_T > 1000 \; \text{GeV}$ test samples respectively (the $p_T$ distributions of the QCD and signal jets generated in this way are very similar, see \App{app:learning}). The third and fourth line in \Eq{eq:80gevprocs} are two examples of the stealth boson $S$ in \Ref{Aguilar-Saavedra:2017zuc}.
The results for the ROC curves giving the signal efficiency versus background rejection are presented in \fig{fig:500_1000_final}.
To better illustrate the effect of the tagging on the signal-to-background significance $S/\sqrt B$, we define significance improvement as the factor multiplying the luminosity-dependent ratio $S/\sqrt B$ due to the tagging, and indicate the lines (in dashed gray) that correspond to a significance improvement of 1, 2, 4 and 8. For comparison we also include the efficiency curve for the dedicate tagger $\tau_{21}^{(1)}$, applied to fat jets from $W$ bosons. Several comments are in order.
\begin{enumerate}
\item The taggers perform better for jets with light quarks, either from $W$ bosons or from stealth bosons decaying to four $u$ quarks.
\item For $W$ bosons the anti-QCD taggers represent a significant improvement over the dedicated tagger $\tau_{21}^{(1)}$, as also observed in \Ref{Datta:2017rhs}.
\item The discriminating power of the {\tt std1000} tagger, when applied to all jet topologies, outperforms the discriminating power of $\tau_{21}^{(1)}$ when applied to $W$ bosons. 
\item Although the discrimination power of the taggers is worse for the two stealth boson examples giving four $b$ quarks, it is far better than the one that would be achieved with  $\tau_{21}^{(1)}$, which is specifically designed for $W$ bosons and actually may reduce the $S/\sqrt{B}$ ratio for this type of signals~\cite{Aguilar-Saavedra:2017zuc}. 
\item The taggers have a good discrimination for fat jets from $H_1^0 \to gg$, for which they are not trained. 
\end{enumerate}
The {\tt std1500} tagger is used to test the performance at higher jet masses and also the ability to distinguish more complex boosted signatures, 
\begin{align}
& H_1^0 \to WW \to q \bar q' q \bar q' \,, && M_{H_1^0} = 400~\text{GeV} \,, \notag \\
& H_1^0 \to gg  \,, && M_{H_1^0} = 400~\text{GeV} \,, \notag  \\
& H_1^0 \to t \bar t \to WbW \bar b, \to q \bar q' b q \bar q' \bar b \,, && M_{H_1^0} = 400~\text{GeV} \,,  \notag \\
& H^\pm \to t \bar b/\bar t b \to Wb \bar b \to q \bar q' b \bar b \,, && M_{H^\pm} = 400~\text{GeV} \,, \notag \\
& H_1^0 \to A^0 A^0 \to b \bar b b \bar b \,,  &&M_{H_1^0} = 400~\text{GeV} \,, \quad M_{A^0} = 80~\text{GeV} \,, \notag \\
& H_1^0 \to Z A^0 \to q \bar q b \bar b \,,  && M_{H_1^0} = 400~\text{GeV} \,, \quad M_{A^0} = 160~\text{GeV} \,,
\end{align}
using a $Z'/W'$ resonance mass of 3300 GeV. These topologies include a $1 \to 1+1 \to 1+3$ asymmetric cascade decay ($tb$), a $1 \to 1+1 \to 2+2$ cascade decay with different intermediate particle masses ($ZA^0$) and even six-pronged fat jets ($t\bar t$) for which the tagger is not trained. The ROC curves are presented in \fig{fig:1500_final}.
As in the previous cases, the discrimination power is best for jets with light quarks, with a significance improvement up to a factor of 8 for $WW$. In addition, it is very good for the rest of signals except for a resonance decaying to $gg$, for which it is not trained. The performance for $t \bar t$ is remarkable, especially if one considers that the tagger is trained with up to four-pronged MI data, and a merged $t \bar t$ jet has six quarks.

The significance improvement from the tagging of shape variables adds to that gained from a jet mass cut. For illustration, we show in \tab{tab:eff1} the significance improvement for the signals 
that is achieved with the cuts $m_J \in [65,105]$ GeV for $\ptj > 500~(1000)$ GeV, as in the {\tt std500} ({\tt std1000}) tagger, and $m_J \in [350,450]$ GeV for $\ptj > 1500$ GeV, as in the {\tt std1500} tagger. The full significance improvement that can be achieved by the combination of jet mass and shape variables is obtained by multiplying the numbers in \tab{tab:eff1} using the ungroomed jet mass with those that can be read from \figs{fig:500_1000_final}{fig:1500_final}. The improvement is modest at low $m_J$ because the ungroomed jet mass distribution for QCD events is large there.

Finally, let us comment about our choice for ungroomed jet masses and $\ptj$. 
There are several methods~\cite{Ellis:2009su,Ellis:2009me,Krohn:2009th,Larkoski:2014wba} to improve signal mass resolution by removing soft particles within the jet. This also tends to improve signal and background separation by shifting the mass spectrum of QCD jets to lower values. However, the results depend on the choice of algorithm and set of parameters, and choices that are optimised for boosted SM particles are often not satisfactory for complex and massive multi-pronged jet topologies such as those considered in this paper. To make this more precise, we compare the significance improvement coming from jet mass cut in \tab{tab:eff1}, between ungroomed jets and jets trimmed~\cite{Krohn:2009th} with the parameters $R_\text{sub} = 0.2$, $f_\text{cut} = 0.05$, commonly used for massive vector bosons $W/Z$. This algorithm and parameter choice work well for $W$ bosons, as can be observed from \tab{tab:eff1}, but is too aggressive for many of the multi-pronged boosted objects for which this groomer can signficantly broaden and shift the signal peak. For example, for stealth bosons with $\ptj > 500$ GeV (row 3, top panel in \tab{tab:eff1}) this groomer degrades the mass resolution. For all the complex signals from $H_1^0$ and $H^\pm$ decays with masses $M=400$ GeV (lower panel, \tab{tab:eff1}), this degradation is more pronounced. Jet pruning~\cite{Ellis:2009me} and soft drop~\cite{Larkoski:2014wba} have a similar performance~\cite{Aguilar-Saavedra:2017zuc}. In any case, the selection of a grooming algorithm and parameter choice that works well for generic BSM objects is another interesting and unrelated issue, which deserves a dedicated study.
\begin{table}[t]
\begin{center}
\begin{tabular}{l|cc|cc}
& \multicolumn{2}{c|}{$\ptj > 500$ GeV} & \multicolumn{2}{c}{$\ptj > 1000$ GeV} \\
& ungroomed & trimmed & ungroomed & trimmed\\ \hline
$W$                                            & 1.54 & 2.13    & 1.52 & 2.24 \\
$H_1^0 \to gg$                           & 1.48 & 1.67    & 1.50 & 1.96 \\
$H_1^0 \to A^0 A^0$ (30) [b]      & 1.48 & 1.46   & 1.46 & 1.83 \\
$H_1^0 \to A^0 A^0$ (15) [b]      & 1.43 & 1.65   & 1.44 & 1.86 \\
$H_1^0 \to A^0 A^0$ (30) [u]      & 1.51 & 1.64   & 1.54 & 1.92 \\
\end{tabular}
\vspace{0.5cm}

\begin{tabular}{l|cc}
&  \multicolumn{2}{c}{$\ptj >1500$ GeV} \\
& ungroomed & trimmed \\ \hline
$H_1^0 \to WW$                        & 2.64 & 2.55  \\
$H_1^0 \to gg$                           & 2.74 & 2.49 \\
$H_1^0 \to t \bar t$                     & 2.43 & 1.33 \\
$H^\pm \to t \bar b / \bar t b$      & 2.37 & 1.99 \\
$H_1^0 \to A^0 A^0$ (80) [b]      & 2.53 & 2.34 \\
$H_1^0 \to Z A^0$ (160)             & 2.39 & 2.07 \\
\end{tabular}

\caption{Enhancement of the significance $S/\sqrt B$ due to the jet mass cuts $m_J \in [65,105]$ GeV (top), $m_J \in [350,450]$ GeV (bottom), with $m_J$ either the ungroomed or trimmed jet mass.
\label{tab:eff1}}
\end{center}
\end{table}

\section{Dedicated versus generic taggers}
\label{app:dedicatedvsgeneric}

\begin{figure}[tb]
\begin{center}
\begin{tabular}{cc}
\includegraphics[height=5.8cm]{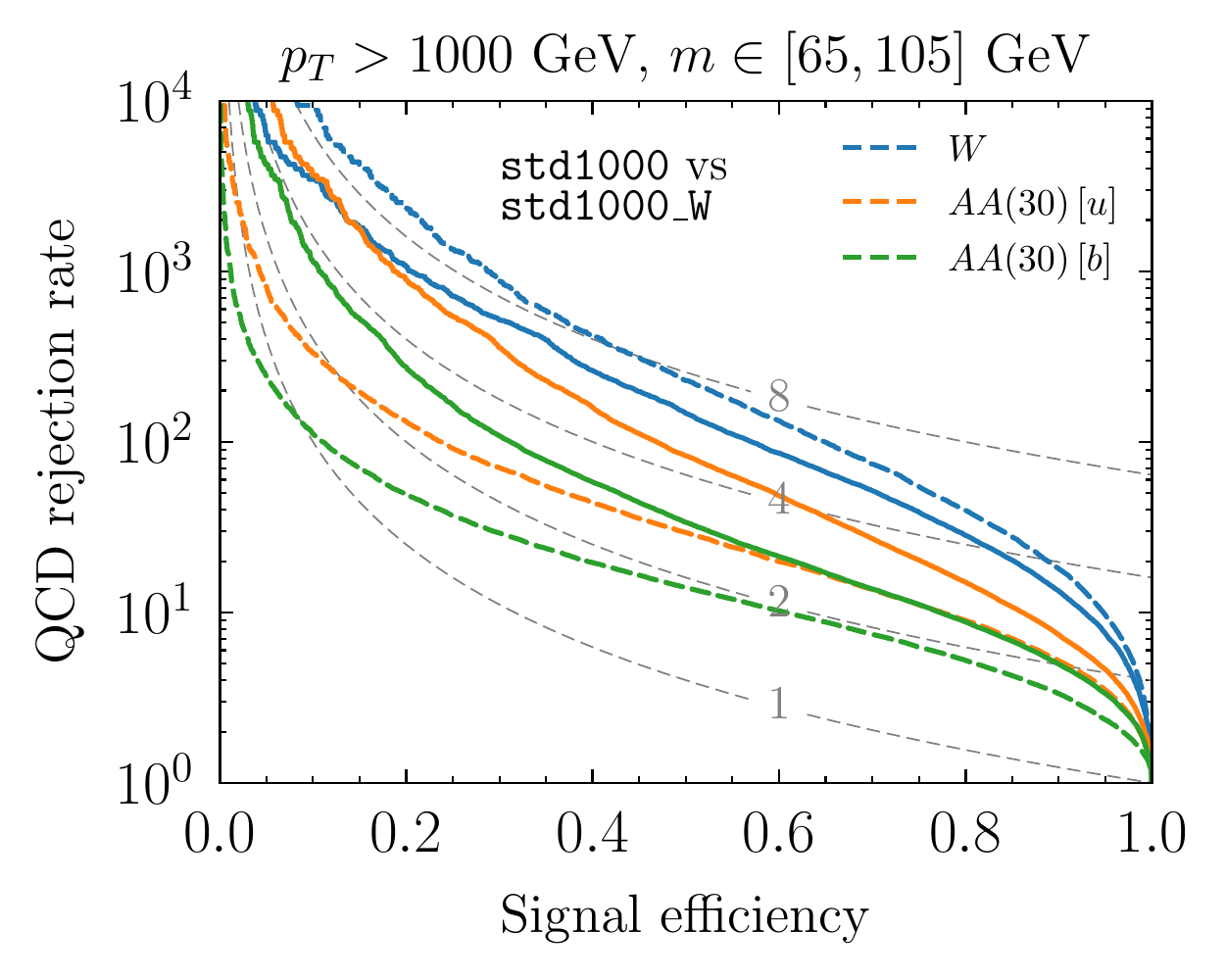} &
\includegraphics[height=5.8cm]{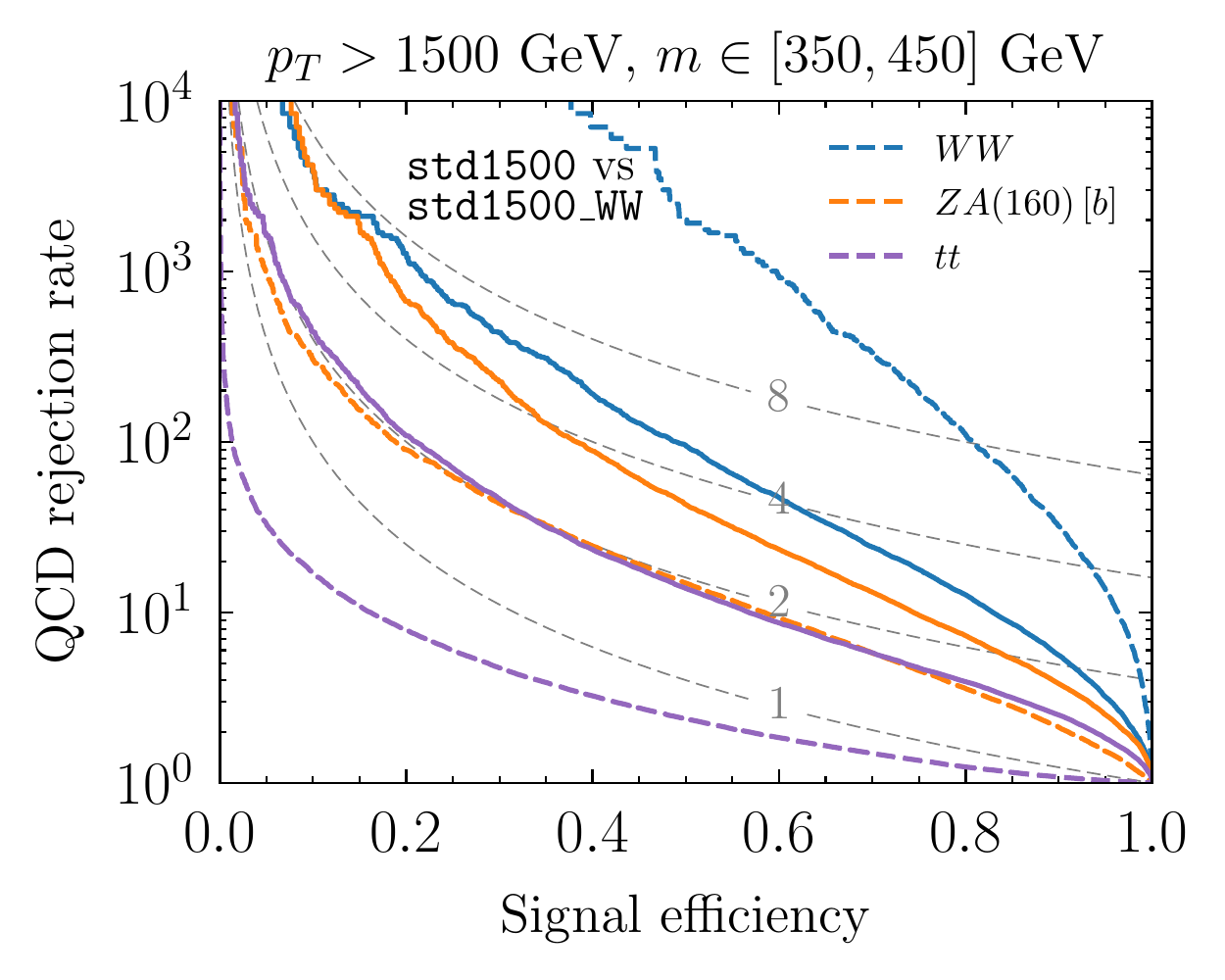}
\end{tabular}
\caption{Performance of dedicated taggers (dashed lines) compared to generic taggers (solid lines). Left: a dedicated $W$-tagger is tested on three signatures at the $W$-mass. Right: a dedicated 400~GeV $WW$-tagger is tested on three 400~GeV signatures.}
\label{fig:genvded}
\end{center}
\end{figure}

It is naturally expected that a NN jet tagger trained on a specific signal will achieve a better discrimination for that signal than a generic tagger, but it will also have a worse performance than the generic tagger on other types of signals. In order to quantify these statements, we have trained two dedicated taggers:
\begin{itemize}
\item[(a)] Tagger `{\tt std1000\_W}': $\ptj > 1000$ GeV, $m_J \in [65-105]$ GeV,  $M_{Z'} = 2200$ GeV, trained on $W \to q \bar q' $ and the QCD background.
\item[(b)] Tagger `{\tt  std1500\_WW}': $\ptj > 1500$ GeV, $m_J \in [350-450]$ GeV,  $M_{Z'} = 3300$ GeV, trained on $H_1^0 \to WW \to q \bar q' q \bar q'$, with $M_{H_1^0} = 400$ GeV and the QCD background.
\end{itemize}
We show our results in \fig{fig:genvded}. On the left panel we can observe that the dedicated tagger {\tt std1000\_W}  has a slightly better discrimination power than the generic tagger {\tt std1000} for the $W$ bosons it is trained on, but somewhat worse for four-pronged stealth bosons. We find that for final states with $W$ bosons, the performance loss by using a generic tagger is rather small, and is more than compensated by the broader sensitivity to new physics signals.
On the right panel it is apparent that the dedicated tagger is significantly better than the generic tagger for the $H_1^0 \to WW$ signal it is trained on, but it is considerably worse for other signals. Firstly, even though this tagger is specifically trained on a four-pronged signal ($WW$), its performance on a different four-pronged signal ($ZA^0$) is even worse than with the generic multi-pronged tagger. This fact illustrates that there can be large differences between signals even if they happen to share the same number of prongs, and justifies our choice to train on MI data rather than specific signal models. Secondly, the sensitivity to $t\bar{t}$ is completely degraded by using the dedicated $WW$ tagger, which actually deteriorates rather than enhances sensitivity to this signal.

\begin{figure}[htb]
\begin{center}
\begin{tabular}{cc}
\includegraphics[height=5.8cm]{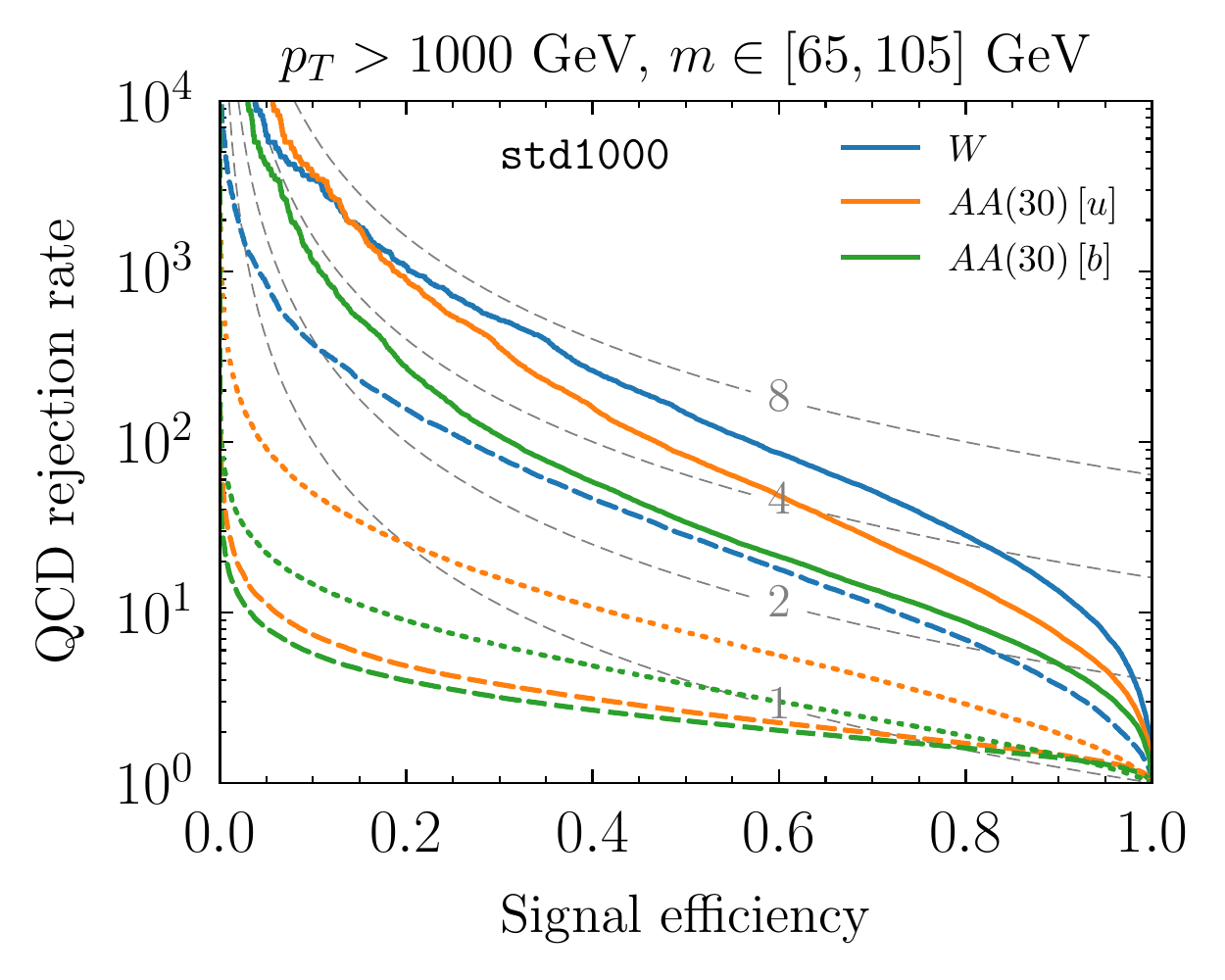} &
\includegraphics[height=5.8cm]{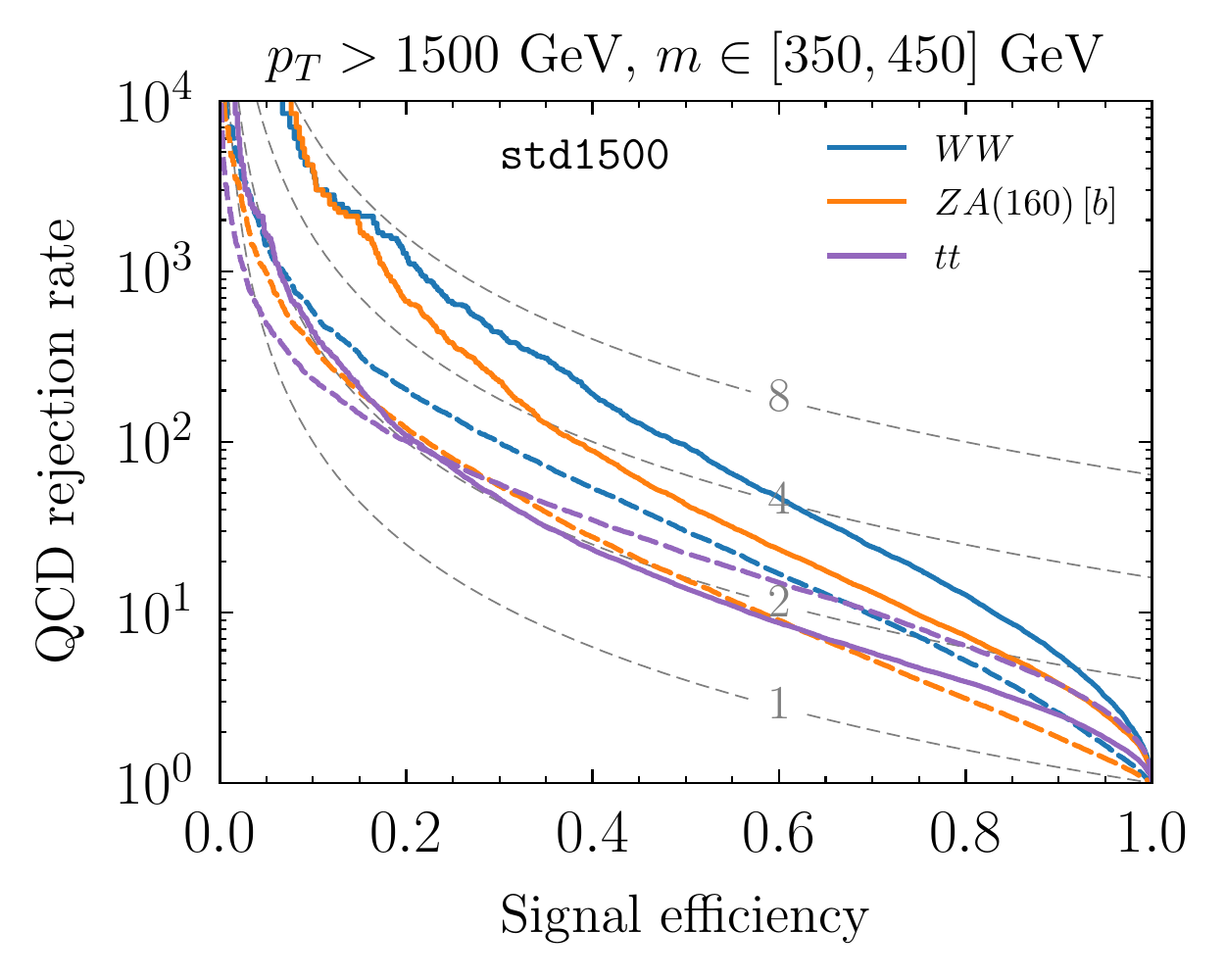}
\end{tabular}
\caption{Comparison of generic taggers (solid) to selected $\tau$-ratios (dashed, dotted), for several boosted signals. Left: The generic tagger \texttt{std1000} is compared to $\tau_{21}^{(1)}$ (dashed) for $W$ and two stealth boson signatures, and also with $\tau_{42}^{(1)}$ (dotted) for the latter. Right: The generic tagger \texttt{std1500} is compared to $\tau_{43}^{(1)}$ for $WW$ and $ZA$, and  $\tau_{63}^{(1)}$ for $t\bar{t}$.}
\label{fig:genvtau}
\end{center}
\end{figure}

Generic multivariate taggers are also found to discriminate the various signals from the QCD background better than the simple $\tau$-ratios that have commonly been used in new physics searches.  In the left panel of \fig{fig:genvtau} we compare the performance of the {\tt std1000} generic tagger (solid lines) to those of $\tau_{MN}^{(1)} = \tau_M^{(1)} / \tau_N^{(1)}$ ratios (dashed) which have been selected for each signal. For $W$-discrimination we compare with $\tau_{21}^{(1)}$ and for four-pronged stealth boson signatures we also use $\tau_{42}^{(1)}$, which has also been used for boosted hadronic $H \to WW^*$ discrimination by the CMS collaboration \cite{Khachatryan:2015bma}. In all the three cases, the performance of the generic tagger is much better, but this is especially apparent for stealth bosons, in agreement with previous results~\cite{Aguilar-Saavedra:2017zuc}. In the right panel we do the comparison for more massive jets using the {\tt std1500} tagger and various selected $\tau$-ratios. Only for a $t \bar t$ signal, for which this tagger is not trained, the performances are comparable.

Altogether, the comparison of generic taggers with dedicated ones and simple $\tau$-ratios is very illustrative. For jet masses around the weak boson masses, there is a remarkable improvement for non-$W$ signals with respect to 
a $W$-dedicated tagger, keeping nearly the same performance for $W$ bosons, and in all cases, quite an improvement over $\tau_{MN}^{(1)}$. For heavier jet masses, the advantage of a generic tagger is still the broader sensitivity, though the performance of a dedicated tagger can be significantly better.


\section{Mass Decorrelation}
\label{sec:4}

\begin{figure}[b]
\centering
\makebox[\textwidth][c]{\includegraphics[width=\textwidth]{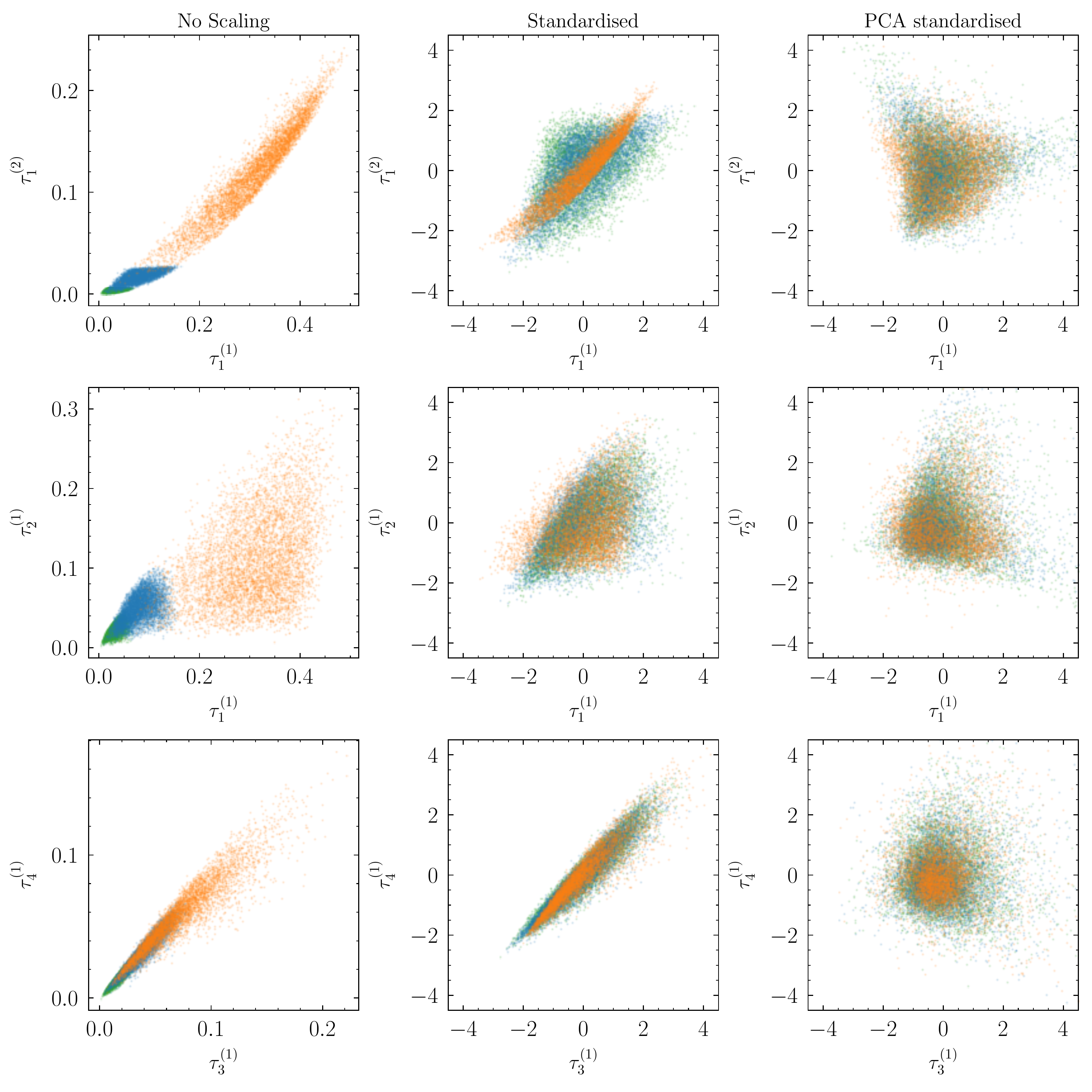}}
\caption{N-subjettiness variables for QCD, plotted in three mass windows and with three different levels of processing. Green: $35 \; \text{GeV} < m_J < 40 \; \text{GeV}$,  blue: $80 \; \text{GeV} < m_J < 85 \; \text{GeV}$,  orange: $265 \; \text{GeV} < m_J < 275 \; \text{GeV}$. Left column: bare $\tau$ variables. Middle column: standardised $\tau$ variables, as described in \Sec{sec:3}. Right column: standardised also along principal component axes. Each row is a different pair of $\tau$ variables.}
\label{fig:taugrid}
\end{figure}

It is desirable, although not compulsory, that a tagger based on jet substructure is decorrelated from the jet mass, in the sense that the tagging efficiency for the background has little dependence on $m_J$. When this happens, the jet tagging does not shape the $m_J$ distribution of the SM background \cite{Dolen:2016kst}.
 This allows for data-driven background estimation using jet mass sidebands, and for the application of bump-hunting strategies on a jet mass distribution. The NNs described in the previous sections, when combined with a choice of threshold on the NN output, act as cuts in the 17-dimensional $\tau_N^{(\beta)}$ space. The distribution of QCD events in $\tau_N^{(\beta)}$-space varies with both jet mass and $p_T$, which is illustrated in the first two columns of \fig{fig:taugrid}. The left column shows a few two-dimensional distributions (there are 136 in total, for the 17 variables considered) in the original $\tau_N^{(\beta)}$ variables, for three intervals of $m_J$. The middle column corresponds to the same distributions, but for the rescaled variables $\tau_N^{\text{std} (\beta)}$ which were used as NN inputs in the previous section.
The efficiency of the tagger (with fixed threshold) on QCD events will necessarily vary with jet mass and $p_T$, and will result in a sculpting of jet mass distributions in ways that depend sensitively on the mass and $p_T$ of the jets on which it was trained.

In order to build a tagger with an efficiency on QCD jets not varyingly strongly with jet mass or $p_T$, there are three obvious possibilities. The simplest one is to apply to the NN output the approach utilised already in~\cite{Sirunyan:2017nvi}. In this case, the threshold on the NN output would be adjusted with jet mass and $p_T$, in such a way that background rejection is fixed. This approach has many advantages (first and foremost being simplicity), but a tagger used in this way that is optimised for signal discrimination at one mass will tend to have suboptimal performance for signals at different masses as the shapes of the input distributions vary, and the tagger might sculpt signal shapes and shift signal mass peaks. It might be required that a suite of taggers are trained, optimised at different mass points. This problem could be ameliorated if a basis of variables is found which are only weakly correlated with jet mass and $p_T$. A second approach, to be adopted in this section, involves preprocessing the $\tau_N^{(\beta)}$ variables in such a way that the QCD distributions of the transformed input variables no longer exhibit strong dependence on mass or $p_T$. This will introduce greater complexity in an experimental analysis which grows with the number of input variables, but will have the advantage that a single tagger can be used with good signal discrimination over a wide range of masses and $p_T$. A third possibility would be to build a tagger that can learn to vary the region of $\tau_N^{(\beta)}$-space to cut as a function of jet mass. This was achieved in \Ref{Shimmin:2017mfk} using an adversarial strategy designed to maintain mass decorrelation on QCD jets. This would leave open the question, however, of how to sample signal masses in training, in such a way that the tagger is not biased towards particular signal masses.

Let us consider a set of $\tau_N^{\text{std} (\beta)}$ calculated using QCD jets selected within a certain jet mass and $p_T$ bin. Arranging the 17 $\tau_N^{\text{std} (\beta)}$ variables into a 17-dimensional vector $\vec{\tau}^{\; \text{std}}$, we define the following transformation for the $\tau_N^{\text{std} (\beta)}$ (in that bin)
\begin{equation}
\vec{\tau}^{\; \text{std}} \to \vec{\tau}^{\; \text{PCA}} = R^{-1} S R \, \vec{\tau}^{\; \text{std}}
\end{equation}
with $R$ and $S$ being 17 by 17 square matrices. $R$ is a rotation matrix that diagonalises the symmetric covariance matrix calculated from this $\tau$-set. This matrix induces a rotation into a basis aligned with the principal component axes of the dataset. In this basis, all pairs of variables are linearly uncorrelated. This is equivalent to choosing a basis whose axes lie along the principal axes of a rigid body formed out of this distribution. We then standardise the data along these axes, so that along each principal axis the standard deviation of the data is 1. This is the action of the diagonal matrix $S$. We then invert the principal axis rotation with the action of $R^{-1}.$\footnote{The reason for applying $R^{-1}$ is as follows. $R$ has a permutation ambiguity, and it is natural to choose this so that the eigenvectors are ordered by the corresponding eigenvalues of the covariance matrix. However, if there is an eigenvalue crossing between two adjacent bins, this will cause a discontinuity in the rescaling matrix $S R$ which would spoil mass decorrelation. The action of $R^{-1}$ removes the permutation ambiguity from the $\vec{\tau}^{\; \text{PCA}}$.} In practice, data should be binned according to jet mass and $p_T$, and a transformation matrix $M_i = R_{i}^{ -1} S_i R_i$ must be determined for each bin $i$. 
Alternatively, one could define the PCA rescaling as a continuous function of $p_T$ and $m_J$ which could be fitted to binned data. In the third column of \fig{fig:taugrid} we plot $\tau^{\text{PCA}}$ distributions for QCD. Firstly, it can be seen that much of the variation in these distributions with jet mass has been eliminated by the transformation. Second, thin directions have been stretched and fat directions have been squashed, as can be seen most clearly in the third row. This fact can aid in the training of the NN.

Therefore, the PCA tagger involves two different tasks:
\begin{enumerate}
\item To set up a transformation map between $\tau$ and $\tau^{\text{PCA}}$, which requires a binning of MC data for the QCD background in $m_J$ and $\ptj$. This map is used both when training the NN (with signal and background events) and when applying the tagger to test data.
\item To train the NN using $\tau^{\text{PCA}}$ variables in some interval of $m_J$ and $\ptj$, which might only be a subset of the entire domain of the transformation map.
\end{enumerate}

\begin{figure}[t]
\centering
 \makebox[\textwidth][c]{\includegraphics[width=15cm]{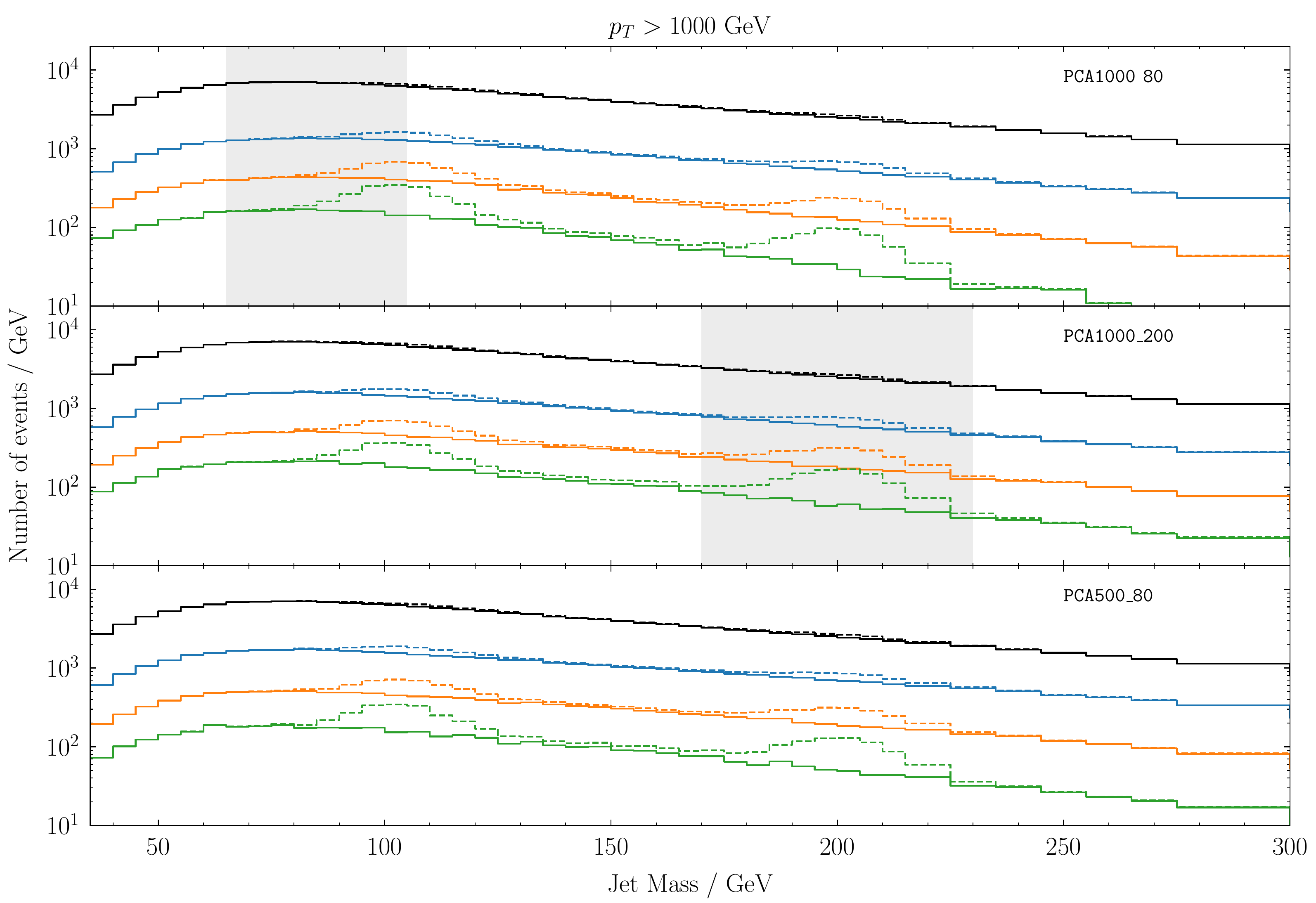}}
\caption{Jet mass distributions for $p_T > 1000 \; \text{GeV}$, selected with three taggers at various thresholds. Top: {\tt PCA1000\_80} tagger. Middle:  {\tt PCA1000\_200} tagger. Bottom: {\tt PCA500} tagger. The solid lines correspond to the QCD jet background, and the dashed lines to the background plus injected signals. The shaded boxes show the jet mass intervals for which each tagger is trained.}
\label{fig:jetmass}
\end{figure}

In order to test whether a tagger trained on input data with this preprocessing will sculpt QCD jet mass distributions, we generate as test data 1,081,834 QCD jets (evenly split between gluon and quark jets) with $p_T > 1000 \; \text{GeV}$, and with no mass cut. The jet mass distribution for this data is given by the solid black lines in \fig{fig:jetmass}. For the PCA preprocessing of the $\tau_N^{(\beta)}$ variables, we bin the data by jet mass with variable bin sizes (as indicated by the bin widths in \fig{fig:jetmass}), in order to have similar numbers of events in each bin, and define a PCA transformation for each bin calculated from this data in that bin.

An additional sample of QCD jets, generated in the same manner as above, is set aside for use in training two new taggers.
 These taggers are trained on the $\tau^{\text{PCA}}$ values of QCD and MI data selected only in a mass window, indicated by the shaded boxes in \fig{fig:jetmass}, to investigate if they will sculpt the QCD jet mass distribution around those windows and if they will still be sensitive to new physics signals outside of those windows. The cuts implemented on the training data and the parameters for the generation of the MI training data for these taggers are
\begin{itemize}
\item[(a)] Tagger `{\tt PCA1000\_80}': $\ptj > 1000$ GeV, $m_J \in [65-105]$ GeV,  $M_{Z'} = 2200$ GeV, $M_S = 80$ GeV.
\item[(b)] Tagger `{\tt  PCA1000\_200}': $\ptj > 1000$ GeV, $m_J \in [170-230]$ GeV,  $M_{Z'} = 2200$ GeV, $M_S = 200$ GeV.
\end{itemize}
The sizes of the event samples used for training these taggers are given in the first two columns of \tab{tab:PCAsamplesizes}. The solid coloured lines in the first two rows of \fig{fig:jetmass} indicate the jet mass distribution for the QCD test sample after selection by the taggers at varying thresholds. We find that there are no new spurious features introduced by application of either tagger.

\begin{table}
\centering
\begin{tabular}{l|l l l}
						& {\tt PCA1000\_80} & {\tt PCA1000\_200} & {\tt PCA500\_80}\\ \hline
Training sample size 	& 108,958       & 71,466	         & 126,372\\
Validation sample size & 27,238         & 17,866          & 31,592
\end{tabular}
\caption{Training sample sizes for {\tt PCA} taggers.}
\label{tab:PCAsamplesizes}
\end{table}

In order to test the sensitivity of these taggers to boosted resonance signals at different masses we simulate the following two signals,
\begin{align}
& H_1^0 \to A^0 A^0 \to b \bar b b \bar b \,, && M_{H_1^0} = 100~\text{GeV} \,, M_{A^0} = 40~\text{GeV} \,, \notag \\
&H_1^0 \to WW \to q \bar q' q \bar q' \,, && M_{H_1^0} = 200~\text{GeV} \,,
\end{align}
resulting from the decay of a 2.2 TeV resonance. The dashed lines in \fig{fig:jetmass} indicate the results when these signals are injected into the QCD test sample, re-weighted to correspond to 1.2\% and 0.7\% of the size of the QCD sample, respectively. We see that both taggers not only succeed in not sculpting the QCD jet mass distribution, but they are also sensitive to BSM boosted objects outside of the mass range in which they were trained.

We also wish to test the effect of using a tagger in a $p_T$ region in which it was not trained. We therefore generate QCD data in the range $500 \; \text{GeV} < p_T < 1000 \; \text{GeV}$, binned in jet mass in the same way as the $p_T > 1000 \; \text{GeV}$ data above. The $\tau$ distributions of this data determine the PCA transformations for data falling into these bins. We also generate MI data on which to train the following tagger,
\begin{itemize}
\item[(c)] Tagger `{\tt PCA500\_80}': $\ptj > 500$ GeV, $m_J \in [65-105]$ GeV, $M_{Z'} = 1100$ GeV, $M_S = 80$ GeV.
\end{itemize}
The size of the event samples used for the training is given in the third column of \tab{tab:PCAsamplesizes}.
This tagger is then applied to the data described above in the $p_T > 1000 \; \text{GeV}$ bins. The results are shown in the third row of \fig{fig:jetmass}. We find that the performance of the tagger is not greatly sensitive to the $p_T$ and mass spectrum of jets used to train the tagger, so long as the data has been properly standardised along the principal component axes.

\section{Discussion}
\label{sec:5}

The generic anti-QCD taggers we have developed in this work provide an alternative to usual taggers in LHC searches for new physics in the boosted regime, with the main advantage being their broad sensitivity to multi-pronged boosted signatures. This feature is of great interest as we do not yet know how new physics might manifest at the LHC. Indeed, new relatively light particles beyond the SM might exist and be produced with very high boosts, for instance if they result from the decay of a heavier particle. If these particles decay hadronically then their signature is a single massive fat jet which might be difficult to separate from QCD backgrounds with existing tools. 

A generic anti-QCD tagger entails a compromise between a high rejection of the QCD background and a broad sensitivity to a variety of signals. As we have shown, a dedicated tagger has a better performance for the specific signal it is trained on, but it can be rather blind to other types of signals. In particular,
\begin{enumerate}
\item For jet masses around the weak boson masses, there is a remarkable improvement for BSM boosted signals (exemplified by stealth bosons) with respect to a $W$-dedicated tagger analogous to the one in \Ref{Datta:2017rhs}, while keeping nearly the same performance for $W$ bosons. Both for $W$ and stealth bosons, the generic tagger provides quite an improvement over the simple ratio $\tau_{21}^{(1)}$ often used in experimental analyses. 
\item For heavier jet masses of a few hundreds of GeV, the advantage of a generic tagger is still the sensitivity to several multi-pronged signals, though the performance of a dedicated tagger can be significantly better.
\end{enumerate}
In either case, final states involving several $b$ quarks are harder to distinguish from the QCD background than those involving light quarks, but $b$ tagging could also be used as an additional independent tool. 
Overall, we observe that searches for new resonances would greatly benefit from a generic tagger for hadronic boosted objects, perhaps complementing dedicated ones. (Dedicated taggers also have their place in specific analyses where one is not interested in other possible signatures, for example in $t \bar t$ measurements in the boosted regime.)

A simple application of a generic tagger of this kind would be an extension of the existing searches for diboson resonances, which search for a resonance bump in a di-jet invariant mass distribution. The use of a generic tagger would allow to search for resonances decaying to a boosted SM boson and a boosted BSM boson. In this case, leptonic decays could be selected for the SM boson and the recoiling fat jet might be selected in a series of broad mass windows after selection by a generic tagger trained in each window. Alternatively, hadronic decays could also be selected for the SM boson, using standard tagging criteria. One recent example is given by \Ref{Aaboud:2017ecz}, which looks for $XH$ decays of a heavy resonance, selecting $H \to b \bar b$ for the Higgs boson and a two-pronged decay $X \to q \bar q\:$ for $X$, with a set of overlapping mass windows for the new particle $X$ and a standard tagger $D_2^{(\beta = 1)}$. In this case, a generic tagger could be used to provide sensitivity not only to $X \to q \bar q\:$ but to other topologies as well. A search could also be carried out for di-BSM bosons, requiring both bosons to have similar mass, and doing a scan over a series of broad mass windows.

Going beyond the discrimination of various signals against the QCD background, it may also be desirable to have a fixed background rejection as a function of the jet mass, for example to allow for data-driven background estimation using jet mass sidebands, and for the application of bump-hunting strategies on a jet mass distribution. This is a solved problem, and can be achieved by applying existing decorrelation techniques to the NN output. However, doing this in such a way as to also maintain good sensitivity to signals over a broad range of masses with a single tagger and without signal-mass bias is a more difficult problem. We have demonstrated that an approach based on standardising along the principal component axes
gives satisfactory results in simulation, which is implemented by building a `transformation map' in the two-dimensional plane of $m_J$ and $\ptj$, using Monte Carlo simulation of the QCD background. This map relates the $N$-subjettiness variables $\tau_N^{(\beta)}$ to the PCA-scaled ones $\tau_N^{\text{PCA}(\beta)}$, which are the inputs to the tagger. This relation varies with the jet mass and $p_T$ and, in practice, it is enough to consider suitable bins in $m_J$ and $\ptj$.  This can be considered as an extension of the approach which has already been taken in a CMS search for light resonances decaying to quark pairs~\cite{Sirunyan:2017nvi} to decorrelate the jet substructure tagger from the jet mass.
In our case, the tagger is trained at some given $m_J$ and $\ptj$ intervals, and it can be subsequently applied outside these intervals by using the map of transformations between $\tau_N^{(\beta)}$ and  $\tau_N^{\text{PCA}(\beta)}$ for other values of $m_J$ and $\ptj$. As \fig{fig:jetmass} demonstrates, this procedure is quite effective. And, in particular, one does not need to train the tagger with various new physics signals at different jet masses and transverse momenta; only the QCD background prediction needs to be known in order to determine the transformations from $\tau_N^{(\beta)}$ to $\tau_N^{\text{PCA}(\beta)}$ at that jet mass and transverse jet momentum.

Although the number of variables ($17 \times 17$ for the correlation matrix) used here for the transformation map of PCA-scaled taggers seems a formidable task for an experimental analysis, let us point out that a simpler approach will suffice. First, a five-body tagger nearly has the same performance, as seen in \App{app:infoinjet}, which reduces the number of variables to $11 \times 11$. Second, some optimisation by reducing the number of variables may be performed too, without sacrificing the performance. Indeed, in this work our goal has been to provide a proof of concept that anti-QCD taggers can be built, leaving the optimisation for future analyses.

Either in its simplest versions (as in \Sec{sec:3}) with standardised input, or in its mass-decorrelated versions (as in \Sec{sec:4}) with PCA scaling, a generic anti-QCD tagger is a novel tool, whose implementation seems feasible, and which could greatly benefit experimental analyses. The final goal is quite ambitious: to enlarge the scope of new physics searches with SM boosted objects, so as to be sensitive to new physics yielding BSM boosted objects. This will constitute a leap forward in new physics searches at the energy frontier, and is well worth the effort.

\section*{Acknowledgements}

We would like to thank Siddharth Mishra-Sharma and Mathieu Cliche for discussions about the implementation of neural networks and python in the early stages of this work. We also thank Rafael Teixeira De Lima, Petar Maksimovic, Nhan Tran, Jesse Thaler and Matt Elsey for discussions and comments on the manuscript. The UFO file used to generate the data used in \App{app:3prongcolour} was generously provided by Bogdan Dobrescu and Felix Yu. The work of JAAS is supported by MINECO Projects FPA 2016-78220-C3-1-P and FPA 2013-47836-C3-2-P (including ERDF), and by Junta de Andaluc\'{\i}a Project FQM-101. The work of JHC and RKM is supported by NSF under Grant No. PHY-1620074 and by the Maryland Center for Fundamental Physics (MCFP).

\appendix


\section{How much information is in a multi-pronged jet?}
\label{app:infoinjet}

\begin{figure}[htb]
\centering
\includegraphics[height=5.8cm]{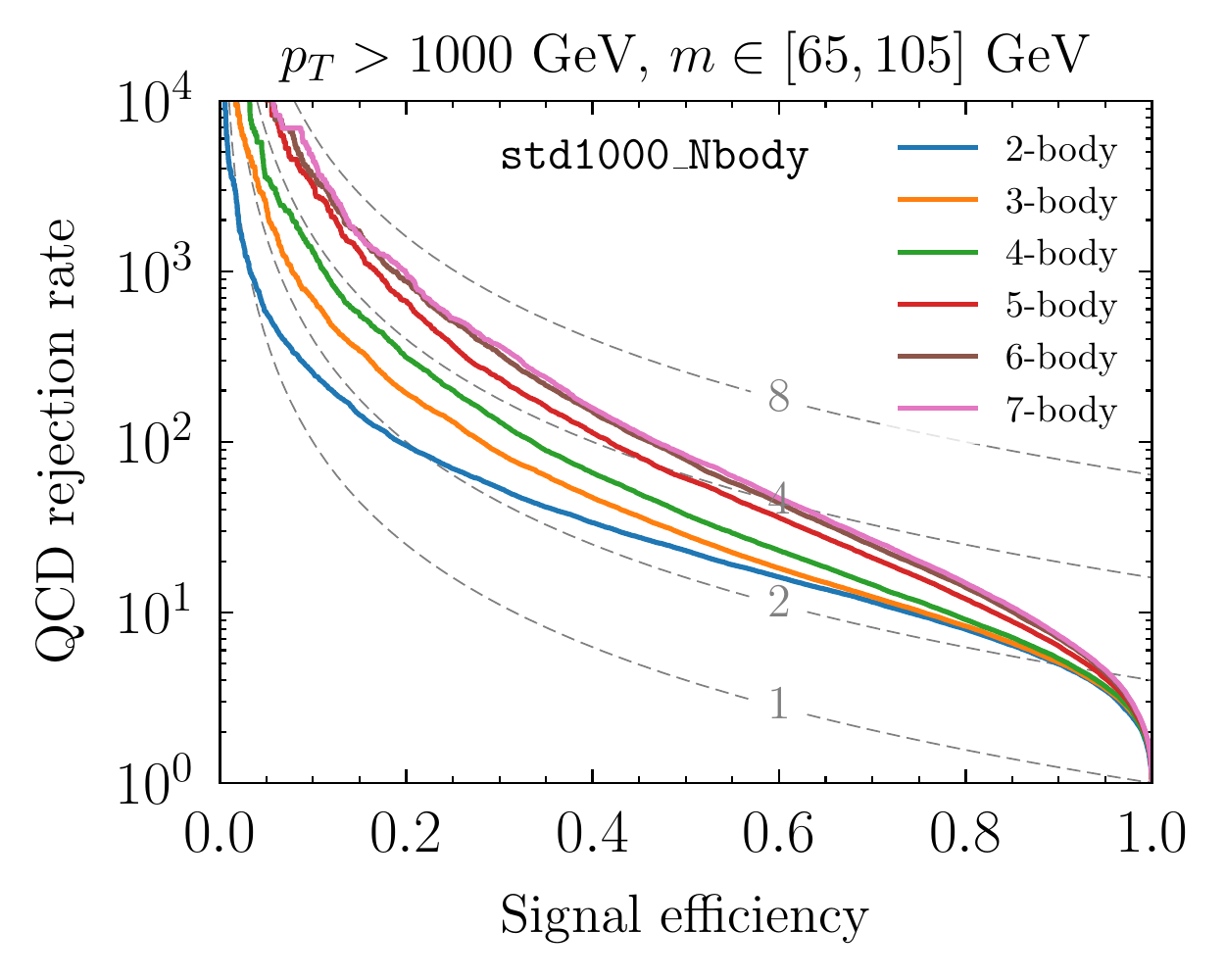}
\caption{Signal efficiency versus background rejection for different versions of the {\tt std1000} tagger taking $M=2,\dots,7$ in  \Eq{ec:tauN}}
\label{fig:Nbody}
\end{figure}

The number of $N$-subjettiness observables that provide additional information on the substructure of multi-pronged jets is found empirically, as in \Ref{Datta:2017rhs}, by training different taggers with $M=2,3,\dots$ in \Eq{ec:tauN} and comparing the ROC lines obtained for the signals of interest. In that reference, for two-pronged jets it was found that the results do not improve beyond $M=4$. Because we also consider three- and four-pronged jets, we have found it is enough to consider $M=7$. For illustration, we show in \fig{fig:Nbody} the performance of different versions of the {\tt std1000} tagger taking $M=2,\dots,7$, applied to a stealth boson signal $H_1^0 \to A^0A^0 \to b \bar b b \bar b$ with $M_{H_1^0} = 80$ GeV, $M_{A^0} = 30$ GeV. For a given $M$, the input to the NN for the corresponding tagger is a $3M-4$ dimensional vector, as defined in~\Ref{Datta:2017rhs}. The ROC curves indicate that the performance saturates before $M=7$.

\section{Effect of signal composition on training}
\label{app:bg-quark-trainOrNot}

The shape of a jet from a heavy quark such as a $b$ quark is in general different from that of gluons and light quarks. We have included light and $b$ quark jets in our MI data earlier, in an attempt to capture all possible shapes, but for simplicity we have not included gluons. In this appendix we show how the results are affected if (i) one doesn't include $b$ quarks in the training data, or (ii) if one also includes gluons. For each case, we train taggers with the MI data set modified -- for case (i) we use the subset of processes in \Eq{ec:MIdata} that do not involve $b$ quark in final state, while for case (ii) we use all the processes in \Eq{ec:MIdata} and in addition we add the process $H_1^0 \to g g$. In all cases, we continue to use equal numbers of events for each of the three or seven categories of training signal data. We perform these studies in two kinematic regimes corresponding to those used for the \texttt{std1000} and \texttt{std1500} taggers in \Sec{sec:3}:
\begin{itemize}
\item [(a)]: $\ptj > 1000$ GeV, $m_J \in [65-105]$ GeV,  $M_{Z'} = 2200$ GeV, and
\item [(b)]: $\ptj > 1500$ GeV, $m_J \in [350-450]$ GeV,  $M_{Z'} = 3300$ GeV,
\end{itemize}
\begin{figure}[t]
\begin{center}
\begin{tabular}{cc}
\includegraphics[height=5.8cm]{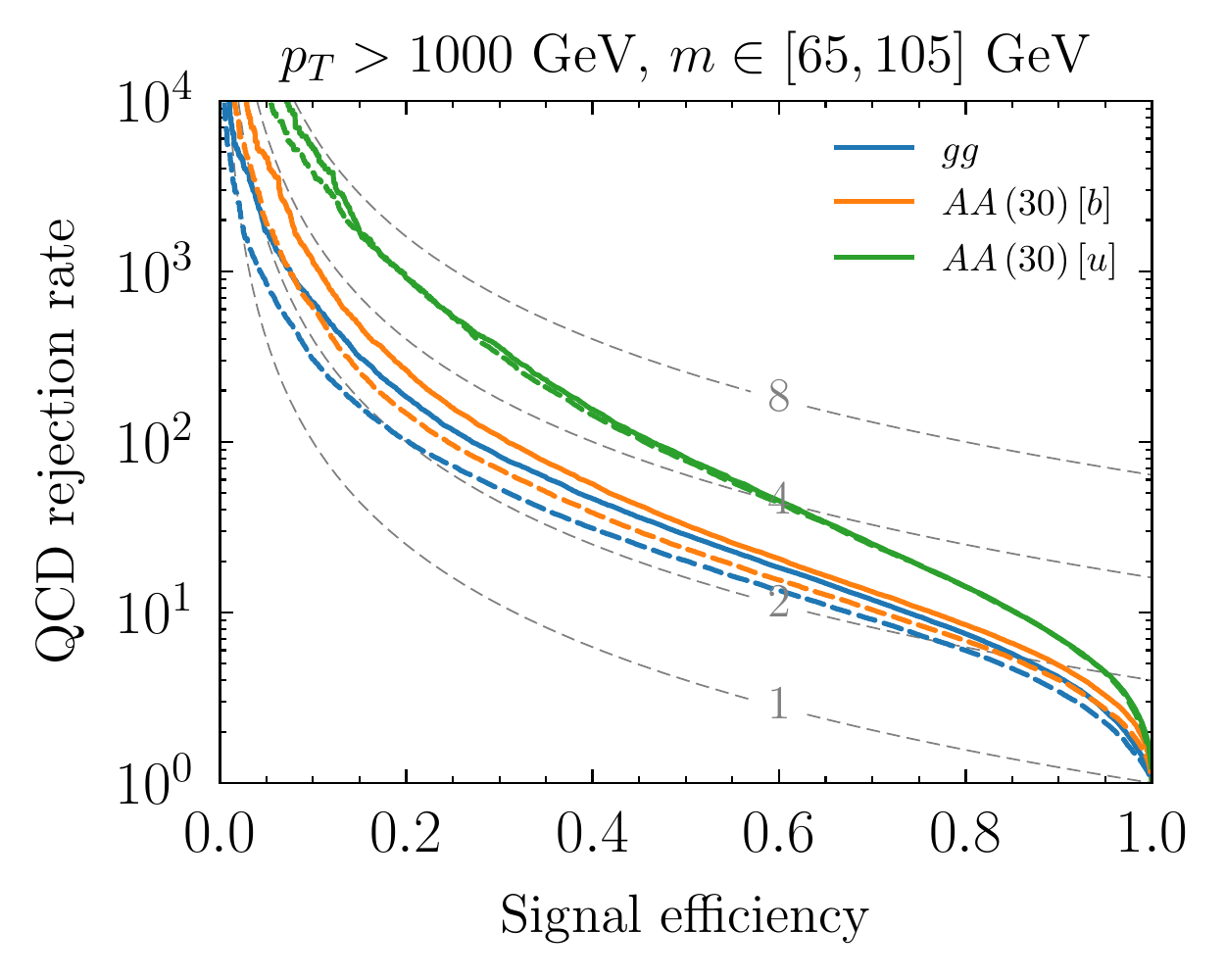} &
\includegraphics[height=5.8cm]{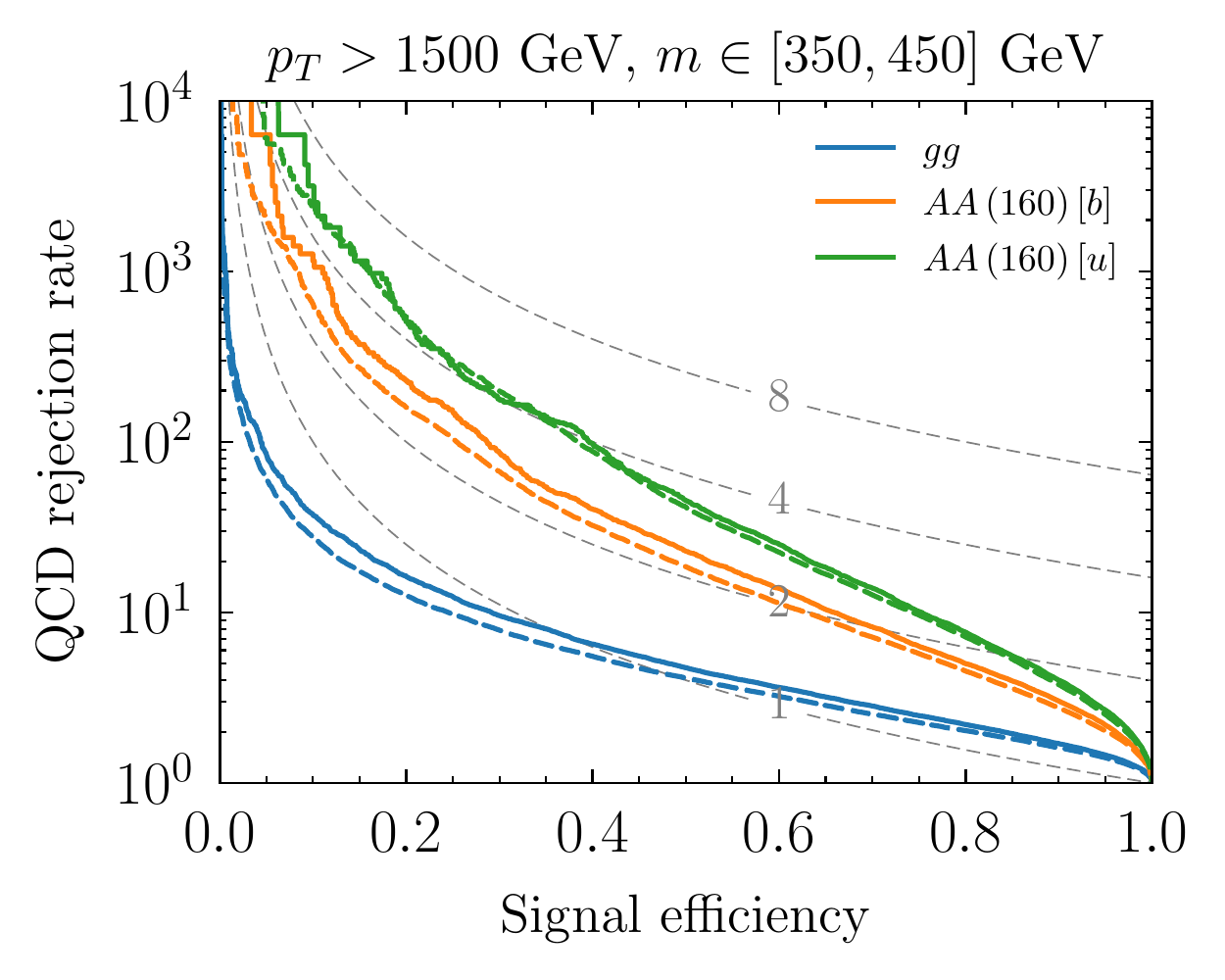} \\
\includegraphics[height=5.8cm]{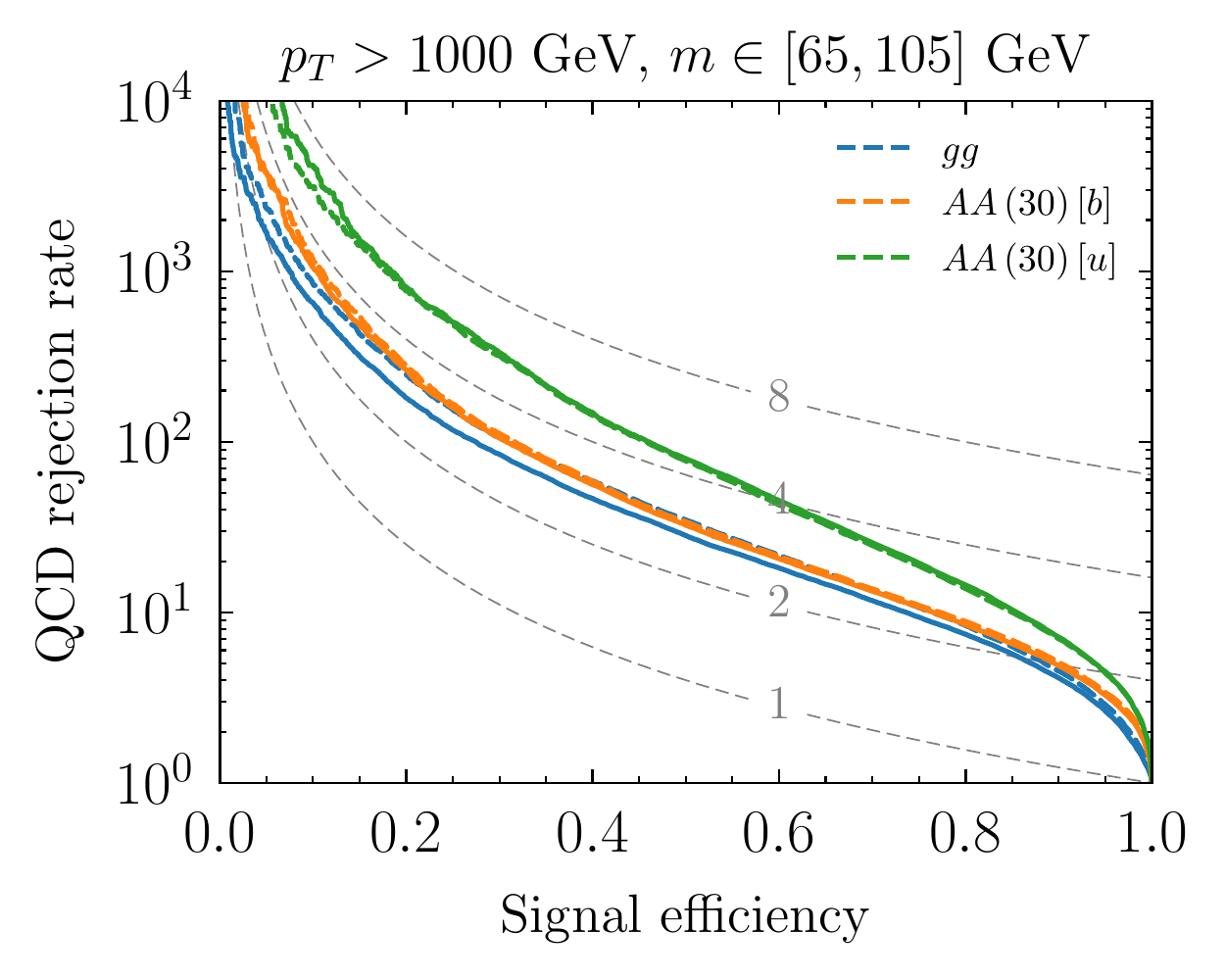} &
\includegraphics[height=5.8cm]{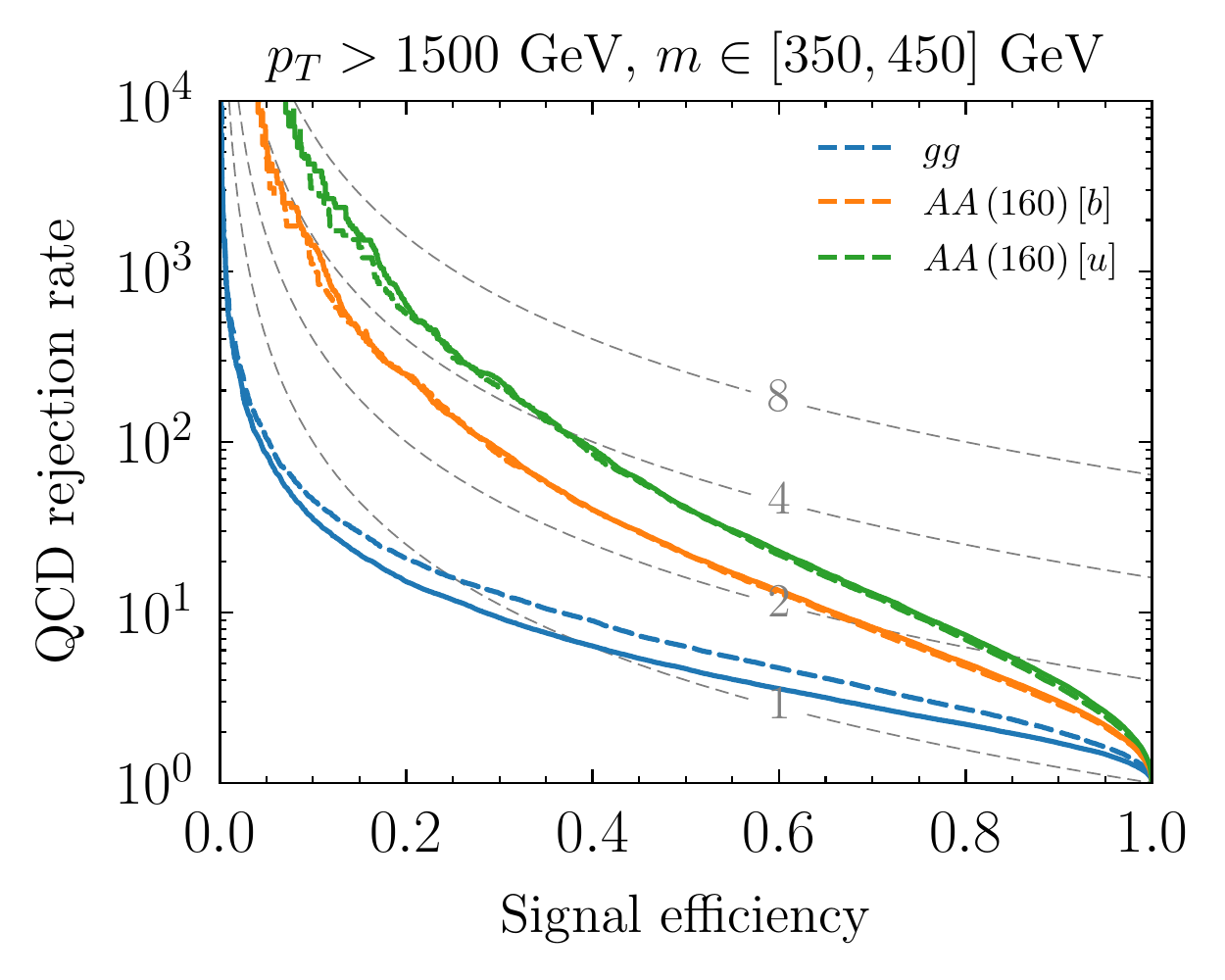}
\end{tabular}
\caption{Top: Effect of $b$ quarks on training, on various test signals. Solid lines correspond to standard choice of MI data for training, and dashed lines correspond to MI data without $b$ quarks in the final state for training. Bottom: Effect of gluons on training, on various test signals. Solid lines correspond to standard choice of MI data for training while dashed lines correspond to MI data as well as gluons in the final state for training.}
\label{fig:BGInTraining}
\end{center}
\end{figure}
%
We test the performance of the taggers on the signal processes
\begin{align}
& H_1^0 \to gg \,, && M_{H_1^0} = 80~(400)~\text{GeV} \,, \notag \\
& H_1^0 \to A^0 A^0 \to b \bar b b \bar b \,, && M_{H_1^0} = 80~(400)~\text{GeV} \,,\; M_{A^0} = 30~(160)~\text{GeV}  \,, \notag \\
&H_1^0 \to A^0 A^0 \to u \bar u u \bar u \,, && M_{H_1^0} = 80~(400)~\text{GeV} \,,\; M_{A^0} = 30~(160)~\text{GeV}  \,.
\end{align}
The masses indicated for $H_1^0$ and $A^0$ correspond to cases (a) and (b) respectively. 
The results are shown in \fig{fig:BGInTraining} (top panel for $b$ quarks and bottom panel for gluons). 

Focussing first on the case of inclusion of $b$ quarks in the training data, we find that for decays with only light ($u$) quarks in the final state inclusion of $b$ quark MI data has no effect on tagging performance. For other decays which include $b$ quarks or gluons in the final state, taggers trained with $b$ quarks do marginally better. Secondly, for the case when $(g g)$-jets are added to the training data, we find that including this process marginally improves the discrimination power for $(gg)$-jets in both kinematic regimes studied. For other processes that have $b$ or $u$ quarks in the final state, the inclusion of these jets in training has a negligible effect on the performance.


\section{Background Composition: Effect of Quark to Gluon ratio}
\label{app:bkgComp}

For simplicity, in the training of our taggers and their testing on Monte Carlo data, we have assumed that the background is composed of equal parts of quarks and gluons. This is obviously not the case in a real analysis, in which the relative ratio will depend not only on the final state considered, but also on the energies involved. In this appendix we explore how sensitive the results are to the precise ratio of quarks and gluons. 

\begin{figure}[h]
\centering
\makebox[\textwidth][c]{\includegraphics[width=\textwidth]{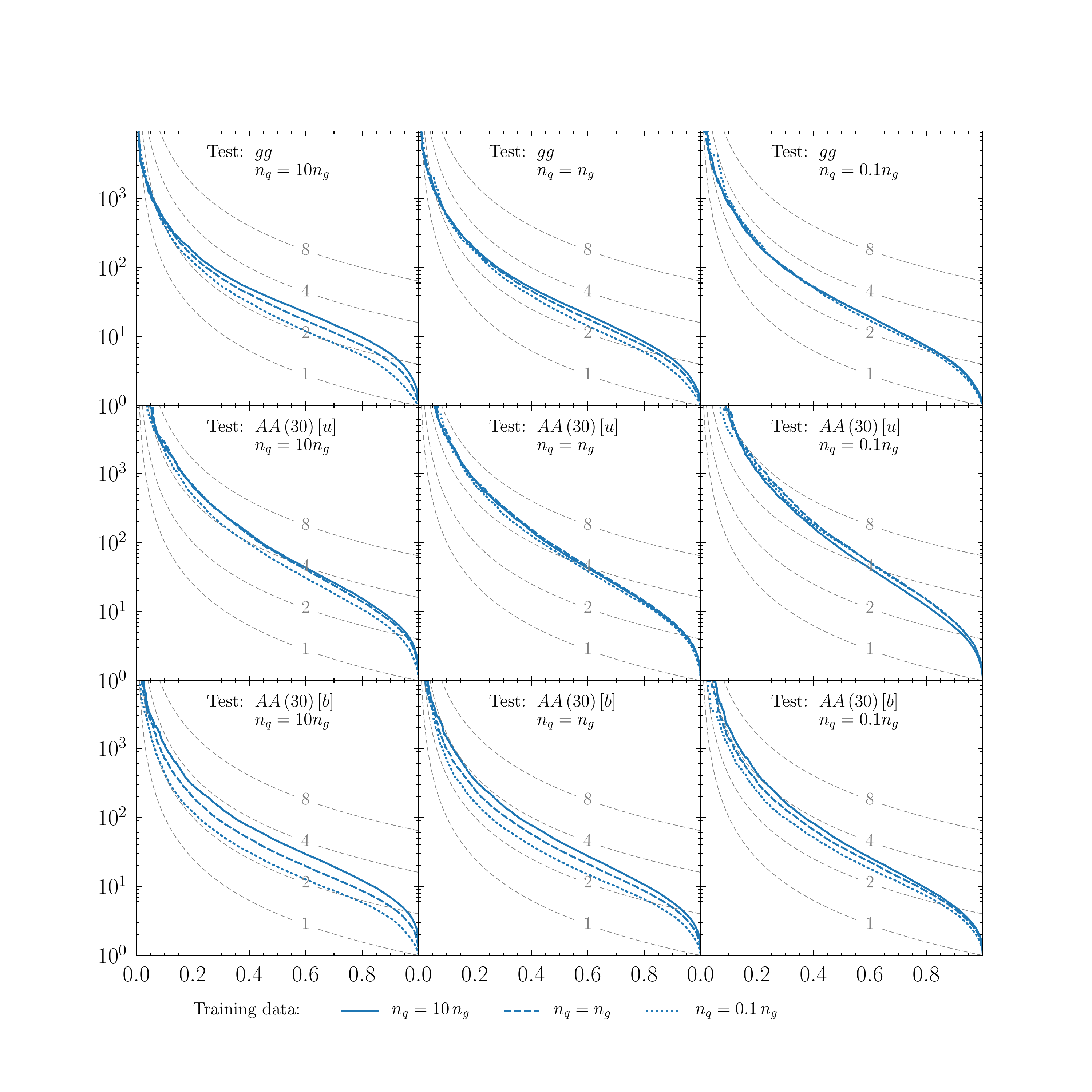}}
\caption{Effect of QCD quark to gluon ratio in training and test data. The performance of the different taggers (solid, dashed and dotted lines) is shown for several signals (in rows) and for different choices of quark to gluon ratios in the background test data (in columns).} 
\label{fig:bkgComp}
\end{figure}

We focus on $\ptj > 1000$ GeV, $m_J \in [65, 105]$, as considered for the {\tt std1000} tagger, and train three taggers on the MI data in \Eq{ec:MIdata} and the QCD background, with three ratios of quarks and gluons:
$n_q = 10 \, n_g$, $n_q = n_g$, $n_q = 0.1 \, n_g$, corresponding to the solid, dashed and dotted lines in \fig{fig:bkgComp}.
We test these taggers for several signals, with a background composed of the same three ratios of quarks and gluons: $n_q = 10 \, n_g$ (left column), $n_q = n_g$ (middle column), $n_q = 0.1 \, n_g$ (right column). The signal processes considered are
\begin{align}
& H_1^0 \to gg \,, && M_{H_1^0} = 80~\text{GeV} \,, \notag \\
&H_1^0 \to A^0 A^0 \to u \bar u u \bar u \,, && M_{H_1^0} = 80~\text{GeV} \,,\; M_{A^0} = 30~\text{GeV}  \,, \notag \\
& H_1^0 \to A^0 A^0 \to b \bar b b \bar b \,, && M_{H_1^0} = 80~\text{GeV} \,,\; M_{A^0} = 30~\text{GeV}  \,, 
\label{eq:c1.test.sig}
\end{align}
The main conclusion of this comparison is that the results actually do not depend much on the precise background composition, as seen from a glance at \fig{fig:bkgComp}. In some cases the relative performance of the three taggers is as expected: for example, for the second process in \eqref{eq:c1.test.sig} above, the tagger is (marginally) better when the background composition is the same in training and testing. But this is not the case for the third process in \eqref{eq:c1.test.sig}. For example, for $n_q = 0.1 \, n_g$, the tagger trained with the `inverse' ratio $n_q = 10 \, n_g$ is slightly better. This suggests that changing the background composition also affects the way in which the tagger learns what is signal and what is background.
\section{NN Architecture}
\label{app:arch}

The choice of architecture in any NN problem merits its own study. Throughout this paper we have used an architecture that gives robust results against variations in its depth (number of nodes in a hidden layer) and breadth (number of hidden layers). In this appendix we show that the results are very insensitive to variations on this choice. Apart from the architecture considered for our results in sections \ref{sec:3}--\ref{sec:4} (two fully connected hidden units with 512 nodes and 32 nodes respectively, henceforth referred as {\tt 512-32}), we consider here two more architectures --- {\tt 1024-32} and {\tt 512-512-32}, in a self explanatory notation.

\begin{figure}[h]
\begin{center}
\begin{tabular}{ccc}
\includegraphics[height=5.8cm]{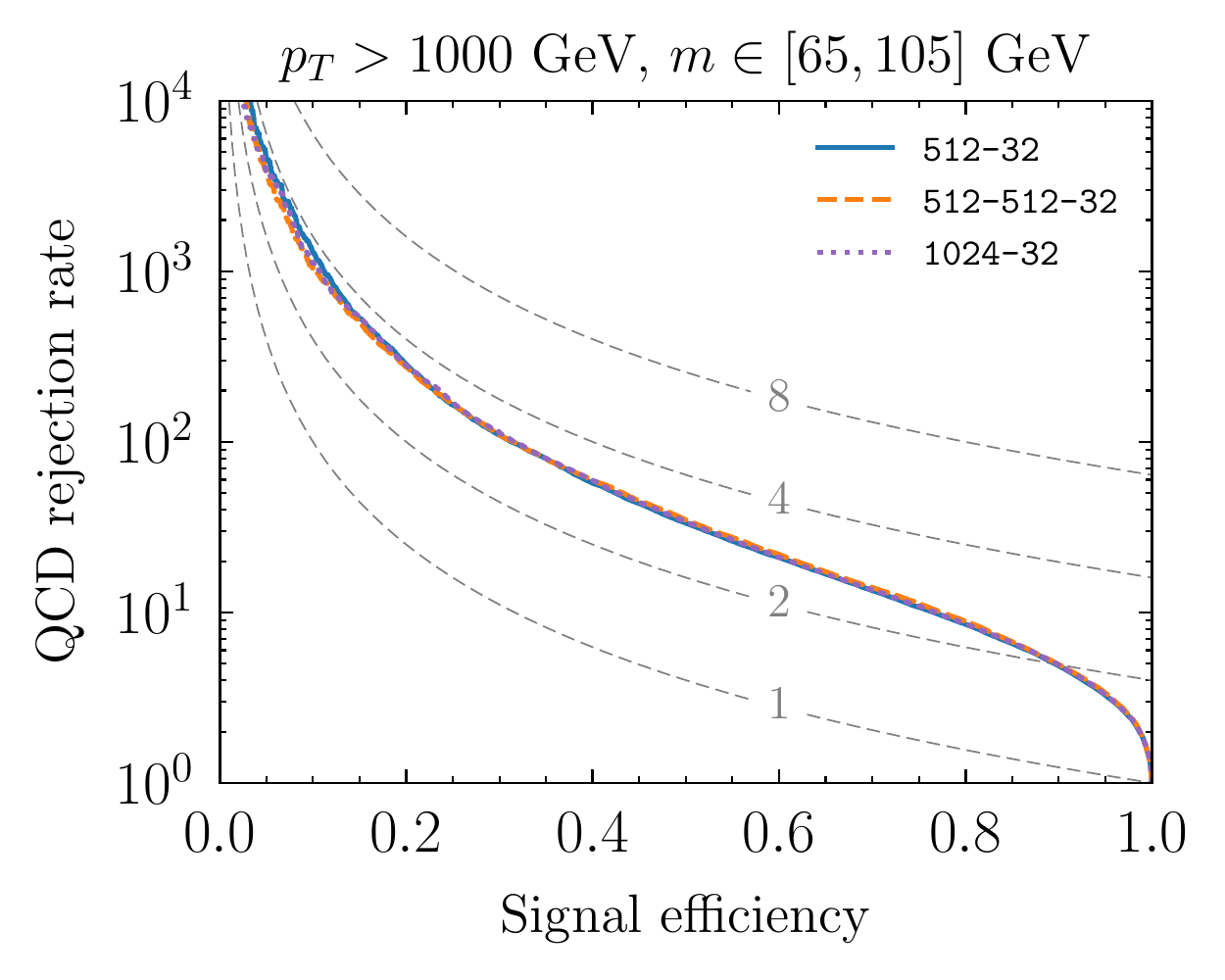} &
\includegraphics[height=5.8cm]{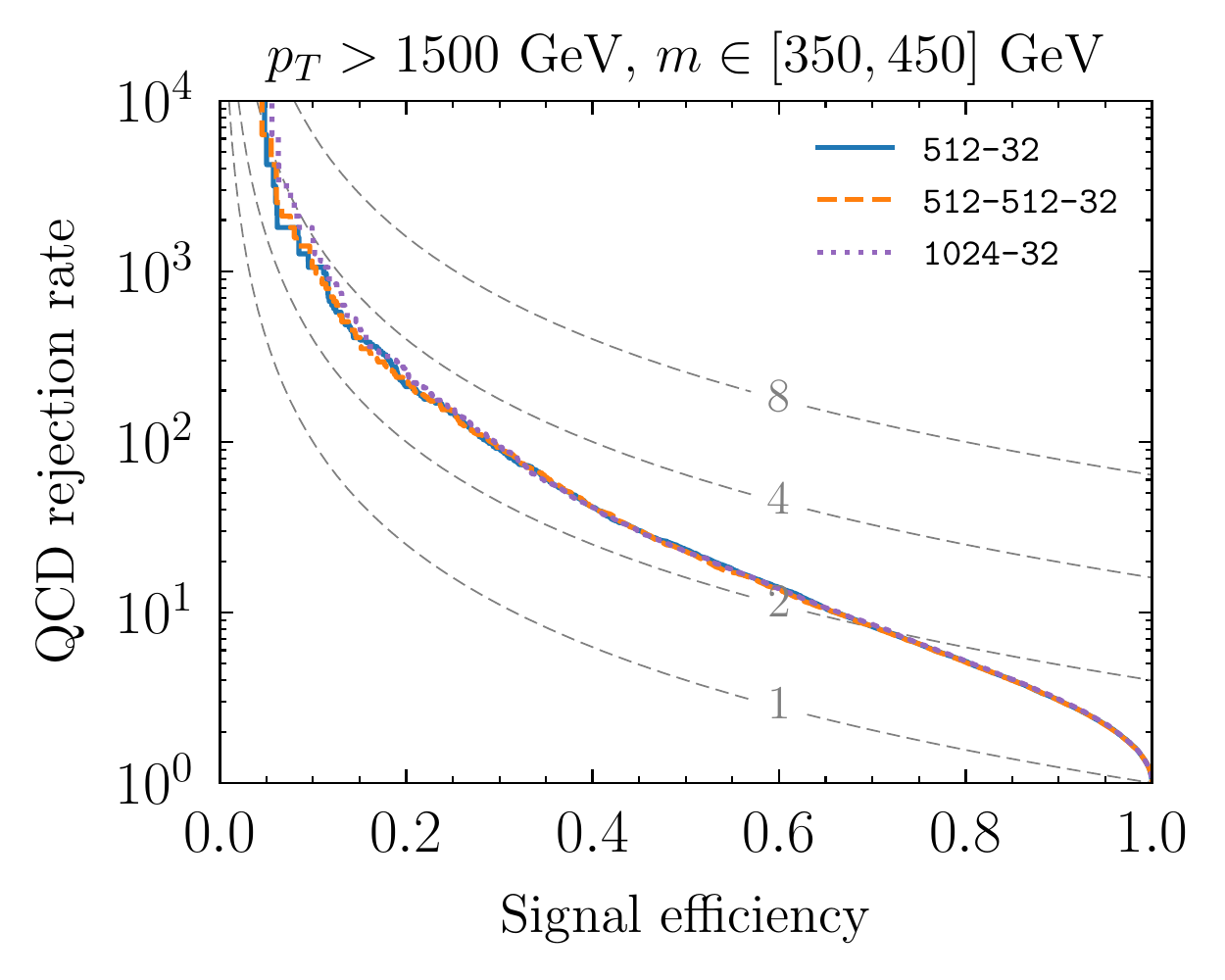} 
\end{tabular}
\caption{Performance of the taggers for different network architectures. }
\label{fig:arch}
\end{center}
\end{figure}

We consider the mass and $\ptj$ ranges used in the definition of the {\tt std1000} and {\tt std1500} taggers, and train two taggers on the MI processes in \Eqs{ec:MIdata}, with $Z'$ masses chosen as in \Sec{sec:3}, and the QCD background. We test the first tagger on stealth bosons with masses $M_{H_1^0} = 80$ GeV, $M_{A^0} = 30$ GeV, and the second tagger with masses $M_{H_1^0} = 400$ GeV, $M_{A^0} = 160$ GeV. The results are shown in \fig{fig:arch}. We see practically no difference in the performance as the NN architecture is varied.

\section{Is the tagger learning shape or kinematics?}
\label{app:learning}

\begin{figure}[t]
\begin{center}
\begin{tabular}{cc}
\includegraphics[height=5.5cm]{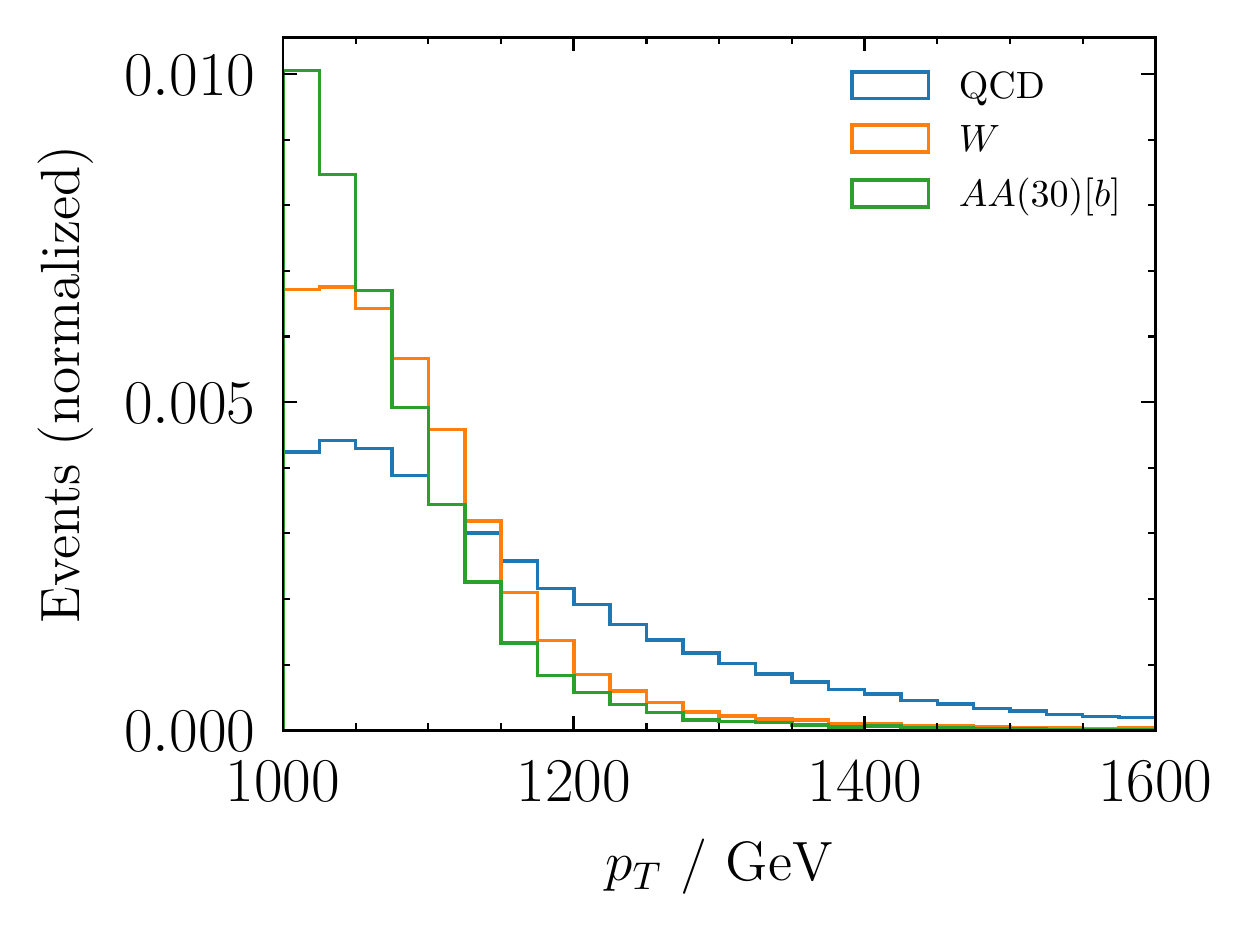} &
\includegraphics[height=5.5cm]{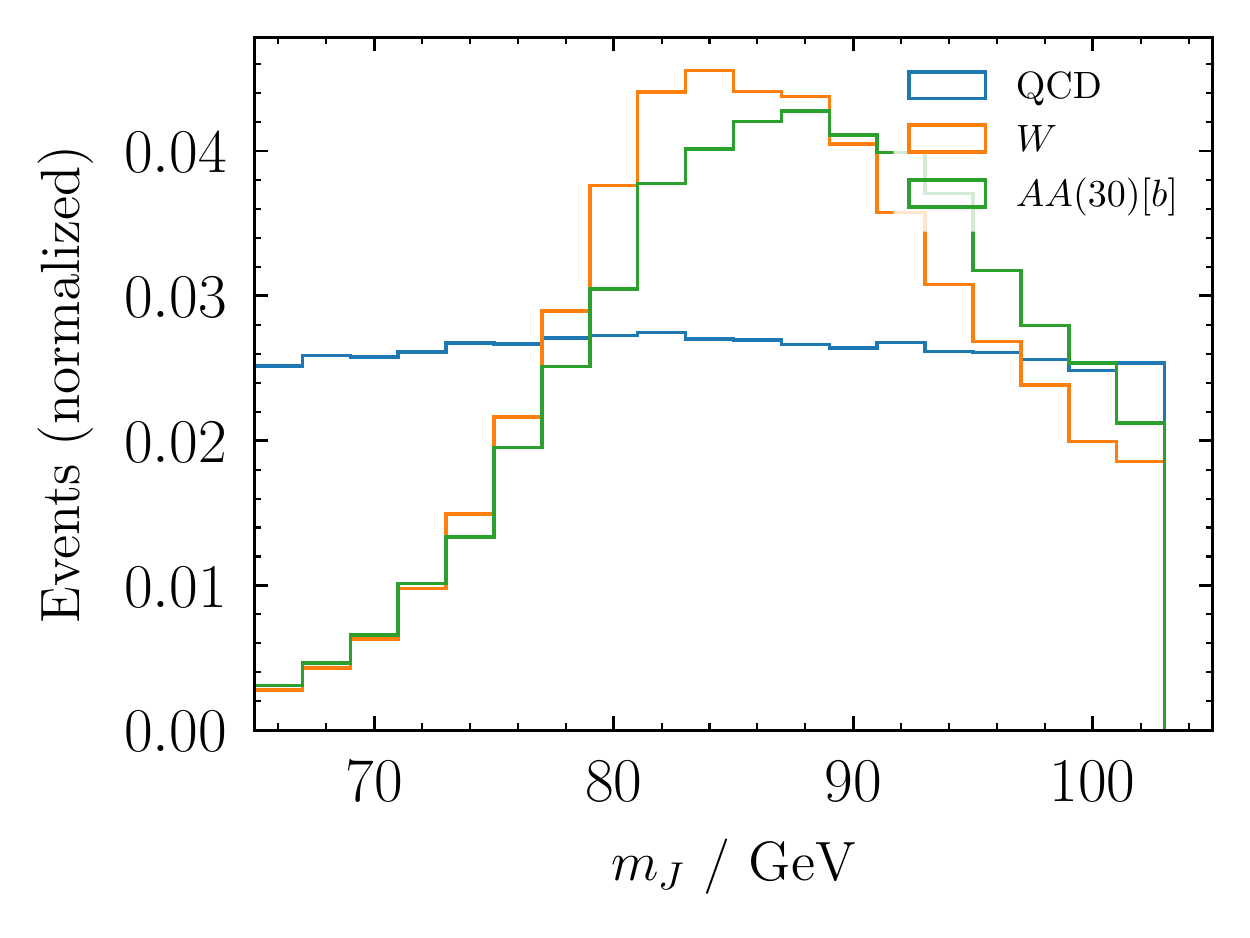}
\end{tabular}
\caption{Transverse momentum (left) and jet mass (right) of the QCD background and two of the signals used to test the {\tt std1000} tagger.}
\label{fig:PTmass1000}
\end{center}
\end{figure}

Although it has been shown that the taggers can efficiently discriminate various multi-pronged signals from the QCD background, a question remains whether this discrimination is solely based on jet shapes or there is also some effect from the different kinematics of the signals and the background. For example, we have already mentioned that the heavy $Z'$ and $W'$ resonance masses have been chosen in such a way that the $\ptj$ distributions are similar to the background, but still there are some differences, which can be seen in \fig{fig:PTmass1000} (left), between the distributions of the QCD background and two sample signals, $W$ and stealth bosons. The same can be said about the jet mass, shown in the right panel: while the background distribution is rather flat, the signals concentrate around and slightly above the input resonance mass.

\begin{figure}[t]
\begin{center}
\begin{tabular}{cc}
\includegraphics[height=5.8cm]{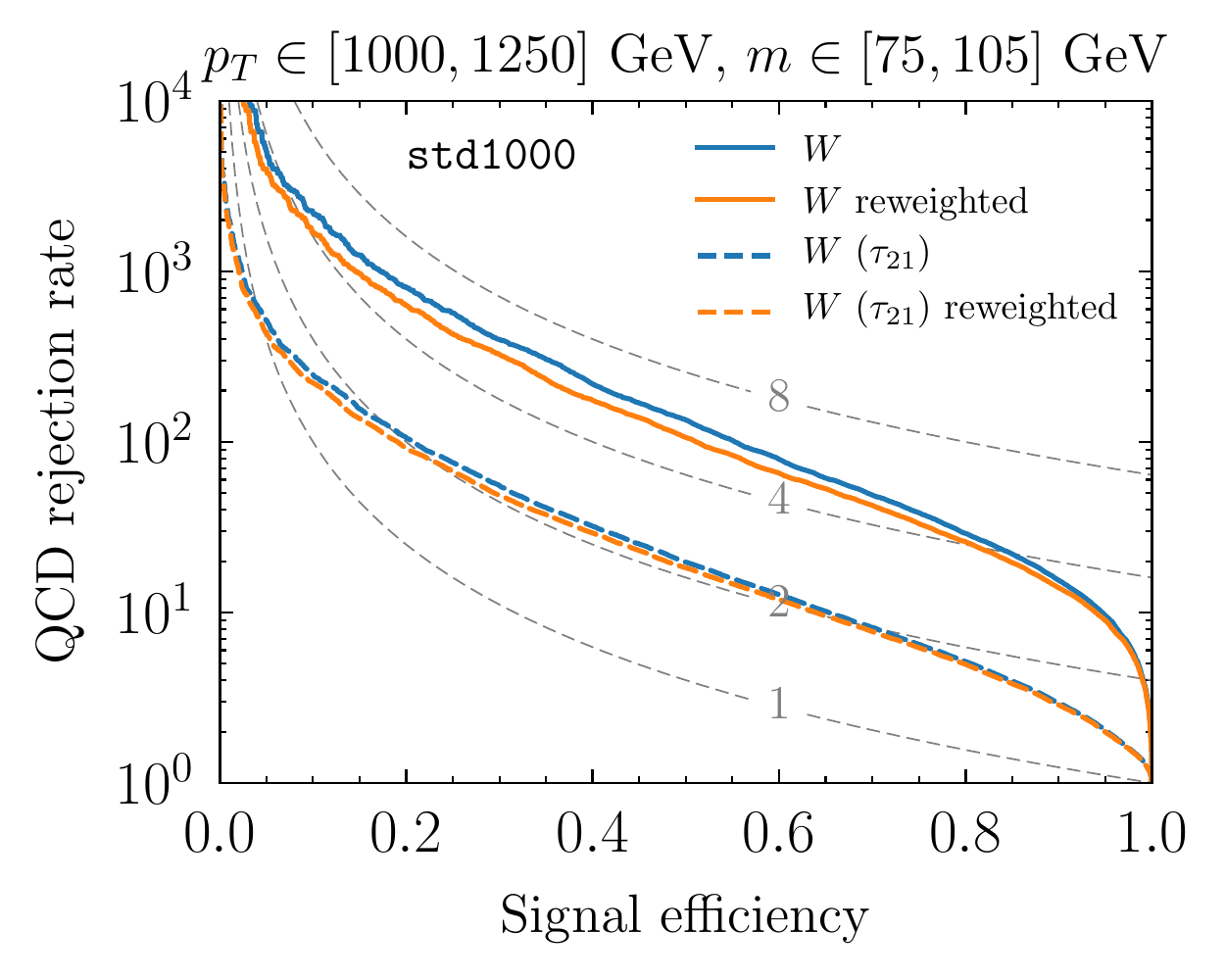} &
\includegraphics[height=5.8cm]{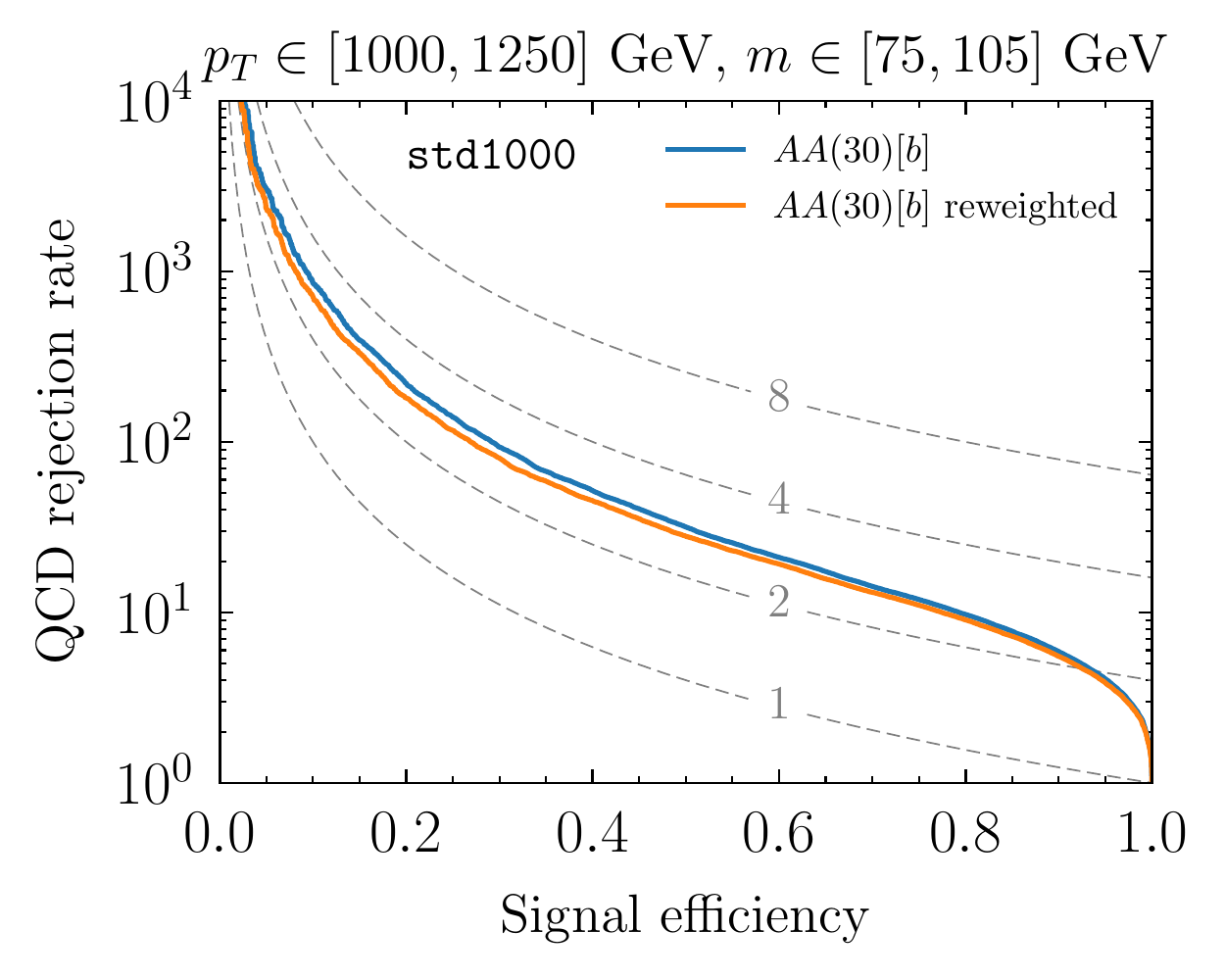}
\end{tabular}
\caption{Comparison between signal efficiency versus background rejection for $W$ boson (left) and stealth boson (right) signals, with their original and with re-weighted $\ptj$ and $m_J$ distributions. }
\label{fig:reweightedroc}
\end{center}
\end{figure}

We have tested the effect of the different $\ptj$ and $m_J$ dependence by considering the discrimination of these two signals with reweighted distributions, using the {\tt std1000} tagger. (The reweighting of the signals makes them have the same two-dimensional $(\ptj,m_J)$ signal distributions as the background.) With this purpose, a two-dimensional binning in $\ptj$ and $m_J$ is applied, with 25 GeV bins in $\ptj$ and 5 GeV bins in $m_J$. The ranges of these variables are restricted to $\ptj \in [1000,1250]$ GeV and $m_J \in [75,105]$ GeV, in order to avoid the appearance of a few events with too large weights that might bias the results. As can be seen from \fig{fig:PTmass1000}, still within those intervals there is significant variation of these two variables.

The comparison between the ROC curves for the signals with the original and re\-weight\-ed distributions, in both cases restricted to the mentioned $\ptj$ and $m_J$ intervals, is presented in \fig{fig:reweightedroc}. The left panel shows 
the results for $W$ bosons, also including the curves for $\tau_{21}^{(1)}$, and the right panel shows the results for stealth bosons. In both cases we observe that the differences between the results with the original and re-weighted distributions are very small. Also, the ROC curves without re-weighting (but with restricted $\ptj$ and $m_J$ range) can be compared to those in \fig{fig:500_1000_final} (right), to see that they are very similar. Overall, it is found that the influence of kinematics in the tagger learning, if any, is quite small. 

\begin{figure}[htb]
\centering
\includegraphics[height=5.8cm]{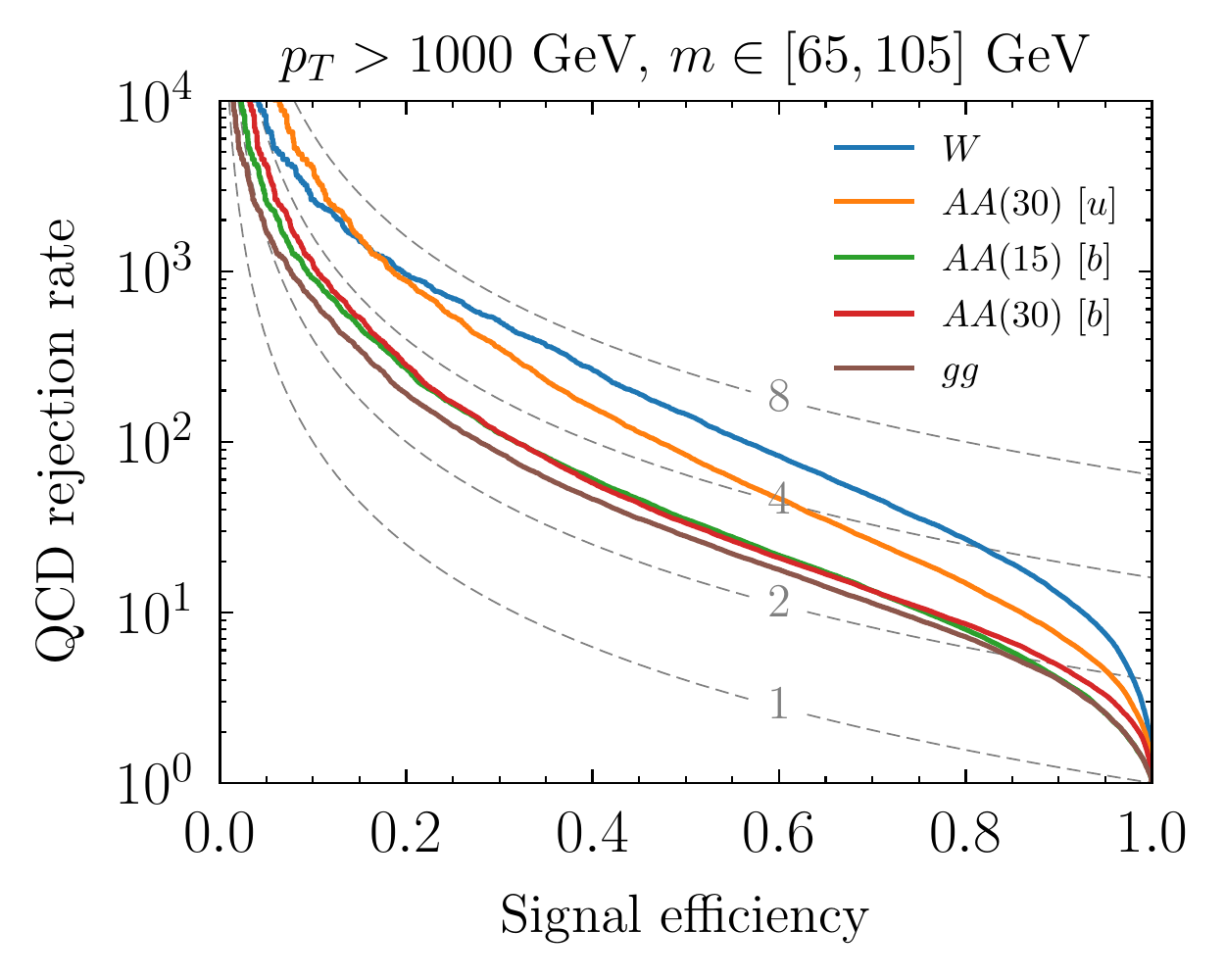}
\caption{Signal efficiency versus background rejection for a variant of the {\tt std1000} tagger which does not use $\tau_1^{(2)}$. The performance is identical to that of the {\tt std1000} tagger in \fig{fig:500_1000_final}.}
\label{fig:no_tau_1_2}
\end{figure}

Besides, we note that the variable $\tau_1^{(2)}$ which is an input to our taggers is closely related to $(m_J / p_T)^2$ \cite{Gras:2017jty}. From the discussions in \Sec{sec:4}, it is our objective to avoid as much as possible jet mass and $p_T$ being directly used as discriminating variables by the tagger. One may wonder whether this variable should have been excluded from our set in \Eq{ec:taulist}. In order to test its influence on our results, we train a variant of the {\tt std1000} tagger on all 7-body variables except $\tau_1^{(2)}$ and compare its performance to the {\tt std1000} tagger in \fig{fig:no_tau_1_2}; we find no effect on the tagger performance. This can be understood because the leading dependence $\tau_1^{(2)} \sim (m_J / p_T)^2$ is the same for all signals and backgrounds at a given mass and $\ptj$. Therefore, the standardisation of the inputs erases this dependence to a very large extent.

\section{Pythia versus Herwig}
\label{app:pythia-herwig}

A serious challenge in the application of machine learning to jet physics in a real collider experiment is the question of whether the distributions of substructure variables are correctly modelled by simulation, and whether the performance of the tagger is robust under mismodelling. Designing approaches to bypass mismodelling fragility is an active area of research~\cite{Dery:2017fap, Cohen:2017exh, Metodiev:2017vrx}, but beyond the scope of this work. In this appendix, we restrict ourselves to investigating the variation of tagging performance when using data hadronised with Pythia (as used for our results in sections~\ref{sec:3}--\ref{sec:4}) and Herwig.

\begin{figure}[h]
\begin{center}
\begin{tabular}{ccc}
\includegraphics[height=5.8cm]{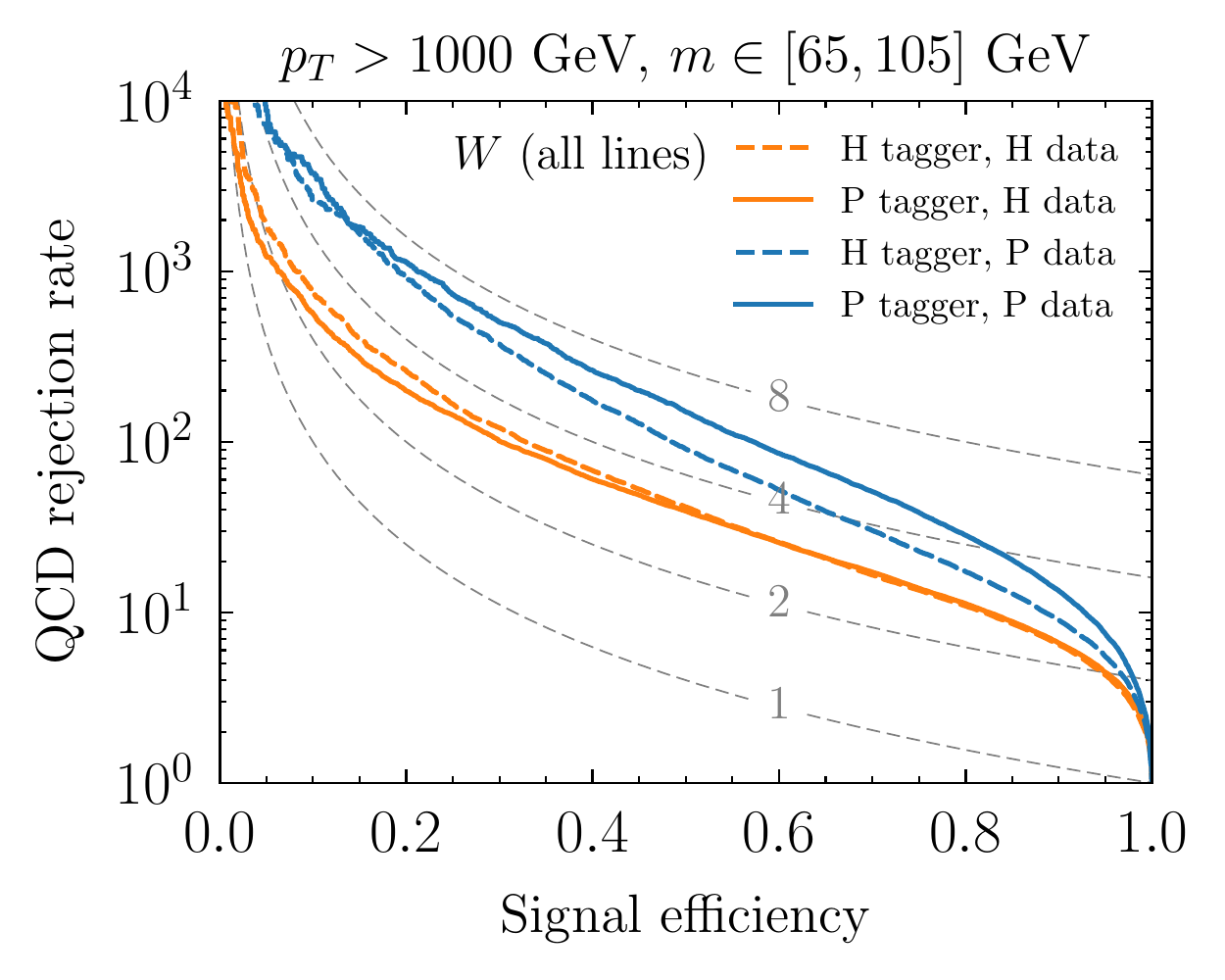} &
\includegraphics[height=5.8cm]{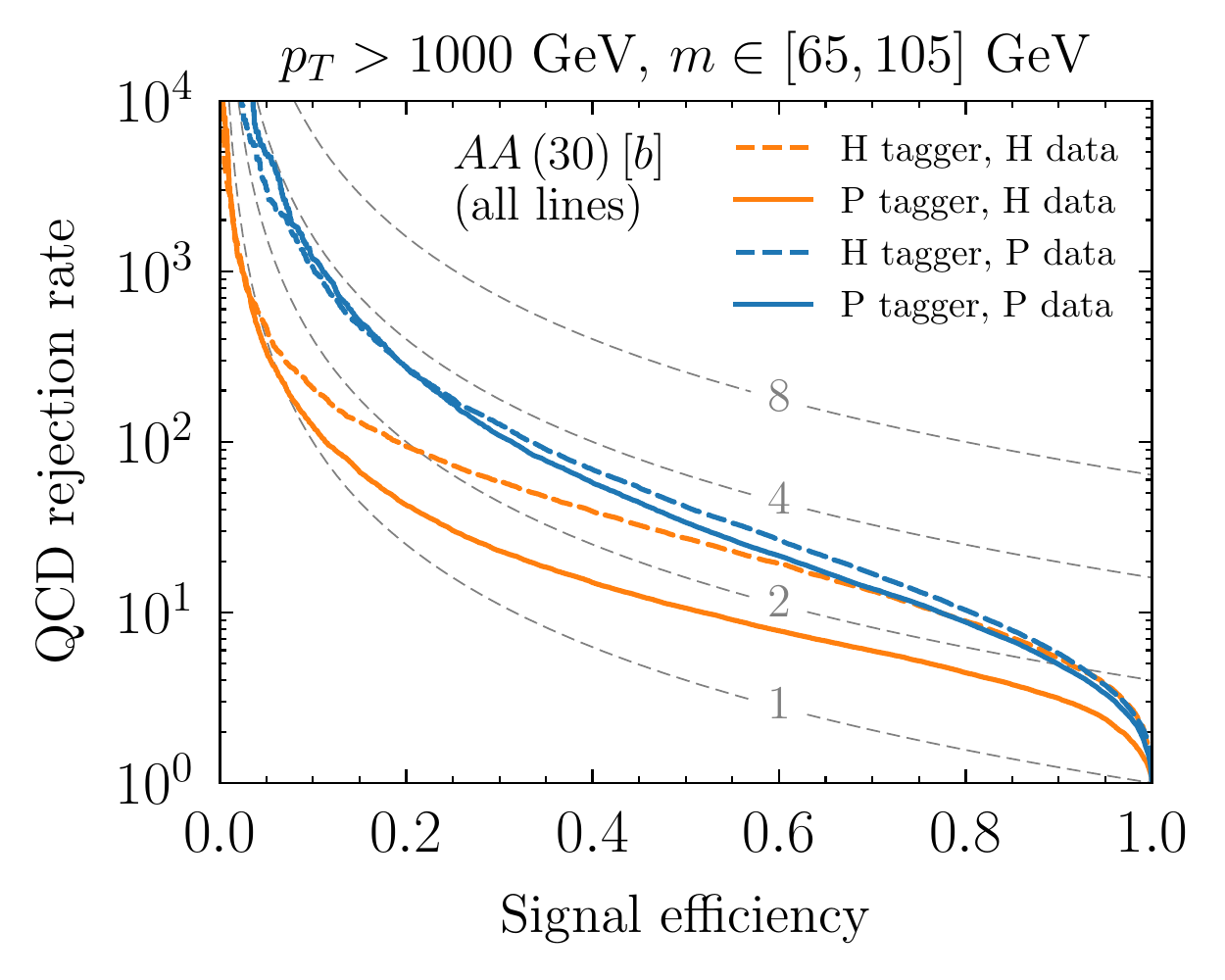} 
\end{tabular}
\caption{Comparison of results using Herwig and Pythia simulated showers, for $W$ bosons (left) and stealth bosons (right).}
\label{fig:pythia-herwig}
\end{center}
\end{figure}

We focus on $\ptj > 1000$ GeV, $m_J \in [65, 105]$, as used in the {\tt std1000} tagger, for brevity. Two new taggers are trained on all processes in \Eqs{ec:MIdata} and the background; one of them is trained on data using Pythia and the other one with data using Herwig. We test the performance of the two taggers on $W$ bosons and stealth bosons with  $M_{H_1^0} = 80$ GeV, $M_{A^0} = 30$ GeV. This test data (both signal and background) is generated twice, once with Pythia and once with Herwig. We show in \fig{fig:pythia-herwig} the results for in-sample tests (e.g. a Pythia trained tagger tested on Pythia data) as well as out of sample tests (e.g. a Pythia trained tagger tested on Herwig data).

We find that in general the performance is better on Pythia generated data than on Herwig generated data, though for the most part this is largely independent of which data the tagger was trained on. The exception is that the tagger trained on Pythia data has significantly worse performance on the Herwig data for stealth bosons, compared to the tagger trained on Herwig data. This difference seems to arise from the different modeling of the signals, as the performance of the Pythia and Herwig taggers on Herwig data for $W$ bosons is quite similar.

Because it is of the utmost importance that the performance of the tagger on QCD data be very well understood, it might be best to train such a tagger with real QCD data, and especially, test its performance directly on data in suitable control regions. Significant uncertainties on signal efficiencies may remain, but these are less important than having an accurate prediction for the background.


\section{Boosted Coloured Jets}
\label{app:3prongcolour}

In this paper we have focused on jets resulting from the boosted decays of colour-singlet new particles, which might easily be missed in searches looking for their direct production due to a small production cross section. Light coloured particles have large production cross sections, and searches for signatures resulting from their direct pair production via QCD are typically highly constraining. However, in~\Ref{Dobrescu:2016pda} it was noted that for some decays of such particles, for example a vector-like quark (VLQ) decaying via a non-renormalisable operator into three light quarks, there are no meaningful LHC constraints from direct searches for these particles with masses between 100~GeV and 1000~GeV. For masses as low as a few hundred GeV, passing the LHC thresholds for jet-based searches may require these VLQs to be produced with high momentum, resulting in collimation of their decay products into a fat jet. Therefore, and also for completeness, it is of interest to see if the generic tagger which we have trained only on colour-singlet jets is sensitive also to these coloured jets. Further, this is the only three-pronged signal which we test our tagger on.

\begin{figure}[htb]
\centering
\includegraphics[height=5.8cm]{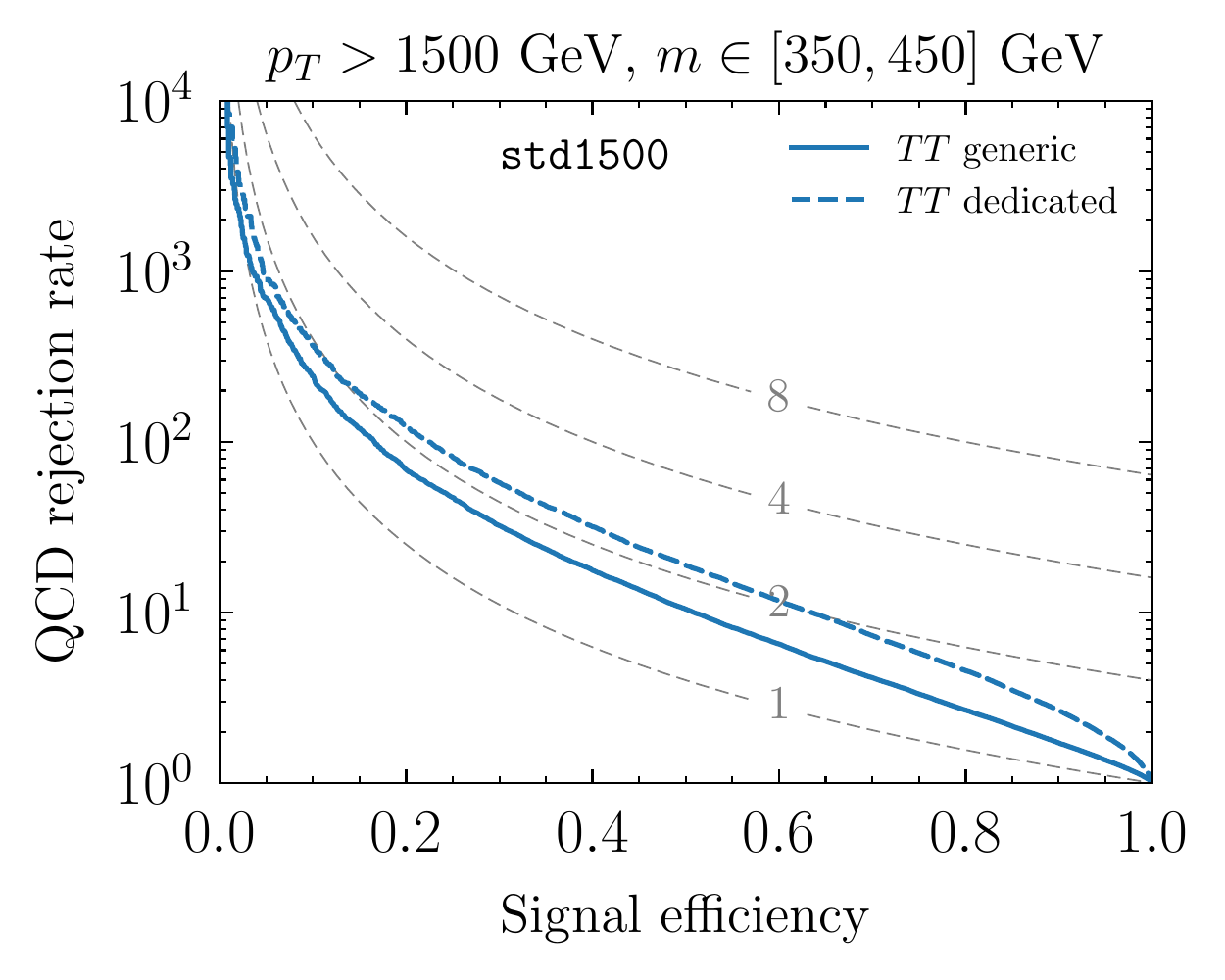}
\caption{Signal efficiency versus QCD rejection for jets resulting from decays of $400 \; \text{GeV}$ vector-like quarks $T$, decaying as $T \to c c \bar{c}$. Solid: generic \texttt{std1500} tagger; dashed: dedicated tagger.}
\label{fig:TTroc}
\end{figure}

We simulate the pair production of VLQs $T\bar{T}$ with $T \to c c \bar{c}$ using an UFO file generously provided by the authors of~\Ref{Dobrescu:2016pda}, setting the $T$ mass to $400 \; \text{GeV}$. We impose a generation level cut $H_T > 3 \; \text{TeV}$, where $H_T$ is the scalar sum of $p_T$ of the quark decay products. The detector level selection is made in the same way as described for the signals used to test the tagger \texttt{std1500} in \Sec{sec:3}, selecting the hardest jet. In \fig{fig:TTroc}, we show the performance of the generic tagger on this signal (solid line), as well as the performance of a dedicated tagger trained to discriminate $T$-jets from QCD-jets. We find that the generic tagger has a moderate performance for this signal, with approximately 10\% QCD efficiency at 50\% signal efficiency. One might wonder if the sensitivity to such signals is lost by training the generic tagger only on colour-singlet jets. However, the dedicated tagger has only marginally better performance, which suggests that the reason for the moderate performance of the generic tagger is that this type of signal is intrinsically hard to distinguish from QCD jets.

An outline of a possible search strategy for this signature could be as follows. Select back-to-back dijet events passing some high $p_T$ threshold. Require the jet mass of the two jets to be close to each other, and after applying a tagger at some threshold, look for a bump in the average jet mass distribution of the two jets. This could either be a cut-and-count analysis in relatively wide jet mass bins, or as a bump hunting shape analysis on a smooth background distribution.

\end{document}